\documentclass[12pt,preprint] {aastex}
\usepackage{epsfig}
\usepackage{psfig}
\usepackage{astro}

\begin{document}
 
\title{THE ANGULAR MOMENTUM OF MAGNETIZED MOLECULAR CLOUD CORES: A 2D-3D COMPARISON} 

\altaffiltext {1} {Service d'Astrophysique, DSM/Irfu, CEA/Saclay, F-91191, Gif-sur-Yvette Cedex, France; sami.dib@cea.fr}
\altaffiltext {2} {Astronomical Institute, University of Utrecht, Princetonplein 5, 3584 CC, Utrecht, The Netherlands}
\altaffiltext {3} {Laboratoire de Radioastronomie, UMR CNRS 8112, \'{E}cole Normale Sup\'{e}rieure, Observatoire de Paris, 24 rue Lhomond, 75231 Paris Cedex 05, France}
\altaffiltext {4} {Harvard-Smithsonian Center for Astrophysics, 60 Garden Street, Cambridge, MA 02138, USA}
\altaffiltext {5} {CNRS/INU, Laboratoire d'Astrophysique de Bordeaux, UMR 5804, BP 89, 33271, Floirac, Cedex, France}
\altaffiltext {6} {Initiative for Innovative Computing, Harvard University, 60 Oxford Street, Cambridge, MA 02138, USA}

\author{Sami Dib\altaffilmark{1,2},  Patrick Hennebelle\altaffilmark{3}, Jaime E. Pineda\altaffilmark{4}, Timea Csengeri\altaffilmark{1}, Sylvain Bontemps\altaffilmark{5}, Edouard Audit\altaffilmark{1}, Alyssa A. Goodman\altaffilmark{4,6}}
 
\begin{abstract} 

In this work, we present a detailed study of the rotational properties of magnetized and self-gravitating dense molecular cloud cores formed in a set of two very high resolution three-dimensional molecular cloud simulations with decaying turbulence. The simulations have been performed using the adaptative mesh refinement code RAMSES with an effective resolution of 4096$^3$ grid cells. One simulation represents a mildly magnetically-supercritical cloud and the other a strongly magnetically-supercritical cloud. We identify dense cores at a number of selected epochs in the simulations at two density thresholds which roughly mimick the excitation densities of the NH$_{3}$ ($J-K$)=(1,1) transition and the N$_{2}$H$^{+}$ (1-0) emission line. A noticeable global difference between the two simulations is the core formation efficiency (CFE) of the high density cores. In the strongly supercritical simulations the CFE is $~33$ percent per unit free-fall time of the cloud ($t_{ff,cl}$), whereas in the mildly supercritical simulations this value goes down to $\sim 6$ percent per unit $t_{ff,cl}$. A comparison of the intrinsic specific angular momentum ($j_{3D}$) distributions of the cores with the specific angular momentum derived using synthetic two-dimensional velocity maps of the cores ($j_{2D}$), shows that the synthetic observations tend to overestimate the true value of the specific angular momentum by a factor of $\sim 8-10$. We find that the distribution of the ratio $j_{3D}/j_{2D}$ of the cores peaks at around $\sim 0.1$. The origin of this discrepancy lies in the fact that contrary to the intrinsic determination of $j$ which sums up the individual gas parcels contributions to the angular momentum, the determination of the specific angular momentum using the standard observational procedure which is based on a measurement on the global velocity gradient under the hypothesis of uniform rotation smoothes out the complex fluctuations present in the three-dimensional velocity field. Our results may well provide a natural explanation for the discrepancy by a factor $\sim 10$ observed between the intrinsic three-dimensional distributions of the specific angular momentum and the corresponding distributions derived in real observations. We suggest that previous and future measurements of the specific angular momentum of dense cores which are based on the measurement of the observed global velocity gradients may need to be reduced by a factor of $\sim 10$ in order to derive a more accurate estimate of the true specific angular momentum in the cores. We also show that the exponent of  the size-specific angular momentum relation are smaller ($\sim 1.4$) in the synthetic observations than their values derived in the three-dimensional space ($\sim 1.8$).   
                                
\end{abstract} 


\section{INTRODUCTION}\label{intro}

An important issue for current ideas of star formation is whether most stars of a given mass are born single, in a binary system or in multiple systems (e.g., Bodenheimer 1995; Bodenheimer et al. 2000; Larson 2010). Among other several physical processes that affect the fragmentation of cores prone to star formation such as their geometry (e.g, Boss 2009), their metallicity (e.g., Hocuk \& Spaans 2010), and degree of magnetization (e.g., Boss 1999; Price \& Bate 2007; Hennebelle \& Teyssier 2008) the rotational properties of the cores will also strongly affect their ability to fragment (e.g., Nelson 1998; Sigalotti \& Klapp 2001; Matsumoto \& Hanawa 2003; Hennebelle et al. 2004; Machida et al. 2009; Walch et al. 2009). This may have important consequences on the fraction of binary and multiple systems. The determination of the angular momentum of the cores and the fraction of their energy stored in rotational motions is complicated by the fact that the velocity fields in the cores usually exhibit complex supersonic motions when mapped with tracers such as CO, CS, and C$^{18}$O with a transition to subsonic, coherent motions when tracers with a higher excitation density are used such as NH$_{3}$ and N$_{2}$H$^{+}$ lines (e.g. Barranco \& Goodman 1998; Goodman et al. 1998; Caselli et al. 2002a). Usually, the assumption is often made in the observations that cores have a uniform rotation and follow a rigid-body rotation law. Their angular velocity $\Omega$ is deduced from measuring global velocity gradients in velocity maps built using velocity measurements along the line of sight. Such measurements for a large number of molecular clouds, clumps or cores were performed by Myers \& Benson (1983), Goldsmith \& Arquilla (1986), Kane \& Clemens (1997), Pound \& Goodman (1997), and Rosolowsky (2007) using CO lines, and by Goodman et al. (1993), Barranco \& Godman (1998) using observations in the ($J-K$)=(1-1) transition of NH$_{3}$, and Caselli et al. (2002a), Pirogov et al. (2003), Olmi et al. (2005) and Chen et al. (2007) using N$_{2}$H$^{+}$ (1-0) line observations. Several other authors have also obtained rotation measurements for individual cores or small numbers of cores (e.g., Harris et al. 1983; Menten et al. 1984; Armstrong et al. 1985; Wadiak et al. 1985; Zheng et al. 1985; Ho \& Haschick 1986; Jackson et al. 1988; Ho et al. 1994; Ohashi et al. 1997; Belloche et al. 2002; Caselli et al. 2002b; Tafalla et al. 2004; Shinnaga et al. 2004; Redman et al. 2004, Schnee et al. 2007; Chen et al. 2008, Chen et al. 2009; Csengeri et al. 2010). 

It is important to investigate whether the distributions of the rotational properties of cores show significant variations as a function of the adopted density tracer/threshold and as a function of the environment (e.g., for example for different magnetization levels of the parent cloud), and most importantly whether the distributions that are determined from observations are a faithful reproduction of the distributions in the intrinsic three-dimensional space, would the entire dynamical and structural properties of the cores be accessible. Furthermore, knowing the statistical distributions of the rotational properties of dense cores in molecular clouds is crucial in order to asses the statistical relevance of rotational parameters assigned to individual rotating core collapse simulations. In previous works, Burkert \& Bodenheimer (2000) superimposed a random velocity field of power spectrum $P(k) \propto k^{n}$ (with $n$ between $-3$ and $-4$) onto the density field of turbulent molecular cloud cores. They found that the projected velocity maps display velocity gradients that can be interpreted as rotation and measured the average specific angular momentum of the cores to be $7 \times10^{20} (R_{c}/0.1)^{1.5}$ cm$^{2}$~s$^{-1}$, where $R_{c}$ is the size of the core, and the average value of the rotational parameter (i.e., ratio of rotational energy to gravitational energy) of $\beta_{rot} \sim 0.03$. Gammie et al. (2003) examined the distribution of specific angular momentum for cores formed in a set of magnetized, self-gravitating molecular clouds with decaying turbulence at the resolution of $256^{3}$. They showed the distribution of specific angular momentum at an intermediate epoch in one of their simulation (the one with beta plasma value of $\beta_{p}=0.1$ after 0.09 sound crossing time). The specific angular momentum they found follows a distribution with values ranging between $10^{21}$ cm$^{2}$ s$^{-1}$ to $7 \times 10^{23}$ cm$^{2}$ s$^{-1}$ with the peak of the distribution at $j=4 \times 10^{22}$ cm$^{2}$ s$^{-1}$. Jappsen \& Klessen (2004) found that there might be a dependence of the average specific angular momentum of the cores as a function of the Mach number of the cloud. Li et al. (2004) and Offner  et al. (2008) also evaluated the specific angular momentum $j$ and rotational parameter $\beta_{rot}$ for cores formed in their molecular cloud simulations, both with resolutions of $512^{3}$ grid cells, and found values $\it {of~the~order}$ of the observed ones. 

We revisit the issue of rotation in molecular clouds cores using very high resolution simulations of magnetized, turbulent, and self-gravitating isothermal molecular clouds (MCs) performed with the Adaptative Mesh Refinment code RAMSES. Our primary aim is to quantify any systematic difference between the intrinsic angular momentum of the cores and the values of the specific angular momentum derived from synthetic observations built using an approach that mimicks the procedure commonly used in the observations. In \S~\ref{simul}, we describe the molecular cloud simulations and the clump finding algorithm used to identify dense cores in the simulations at different density thresholds. In \S~\ref{general} we describe some of the basic properties of the simulations. The rotational properties of dense cores are presented and discussed in \S~\ref{comp} and in \S~\ref{conc} we summarize our results and conclude.

\section{SIMULATIONS AND CLUMP-FINDING}\label{simul}

In this study, we analyze two three-dimensional, isothermal, magnetized, and self-gravitating molecular cloud simulations with decaying turbulence. The simulations were performed using the adaptative mesh refinement (AMR) code RAMSES (Teyssier 2002) on a cubic grid with periodic boundary conditions in all directions. RAMSES is a second order Godunov scheme and uses the constraint transport method to ensure ${\rm div} {\bf B}=0$ (Fromang et al. 2006). The initial setup of the simulations is similar to those presented by Dib et al. (2007a,2007b,2008a,2008b) on uniform grids with $256^{3}$ and $512^{3}$ resolutions. However, by using the AMR technique, the linear resolution is enhanced by almost an order of magnitude which is extremely important in order to study the properties of dense cores which occupy but a small fraction of the simulation volume. The initial coarse grid is uniform and consists of $512^{3}$ cells. The density field on the initial grid is homogeneous. Refinement of the grid was performed such as to keep a local Jeans mass resolved with 10 grid cells. The grids were refined for up to three additional levels leading to an effective resolution of $4096^{3}$ grid cells\footnote{However, at the last time-steps included in our analysis, the number of grid cells in the most refined level remains quite modest and is of the order of a few thousand cells (e.g., 5311 cells at the last considered time-step in simulation B2).}. In contrast to our previous simulations in which turbulence was continuously forced, in the current simulations, turbulent motions that are injected at the start of the simulations are left to decay freely as time evolves. The initial local components of the turbulent velocity field are modeled using a fractal brownian motion power spectrum with an exponent of -11/3, corresponding to a Kolmogorov spectrum. The initial velocity perturbations are generated on all scales of a $256^{3}$ cells grid and interpolations are performed to acquire the velocity components on the initial coarse $512^{3}$ grid. The amplitudes of the velocity components are chosen such as to represent an initial $rms$ sonic Mach number\footnote{The initial Alfv\'{e}n Mach number in simulations B1 and B2 is thus equal to ${\cal M}_{A} = {\cal M} \times \beta_{p}^{1/2} = 0.1$ and $10$, respectively} of ${\cal M}=10$. The Jeans number of the box is $J_{box}=4$ (i.e., number of Jeans masses in the box is $M_{box}/M_{Jeans,box}=J_{box}^{3}=64$). 

In physical units, the simulation box has a size of $L_{cl}=4$ pc (thus the spatial resolution is $4~{\rm pc}/4096 \sim 10^{-3}$ pc), an average number density of $n_{aver}=500$ cm$^{-3}$, and a temperature of $T=11.43$ K which leads to a sound speed of $0.2$ km s$^{-1}$ with a mean molecular weight of $M_{w}=2.36$ in units of the proton mass $m_{H}$. The free-fall time of the simulation volume is defined as being $t_{ff,cl}=1/(G~\rho)^{1/2} \sim 2.76 \times 10^{6}$ yrs, where $\rho$ is the physical uniform density in the box\footnote{The value of $t_{ff,cl}$ defined here is larger than the free-fall time of a sphere of uniform density $\rho$ by the factor $(32/3~\pi)^{1/2}$}. The simulations differ by the strength of the initial magnetic field in the parent cloud, with one simulation being slightly magnetically-supercritical (run B1), and the other one strongly supercritical (run B2). Initially the magnetic field has a component only in one direction. The initial magnetic field strength, the mass-to-magnetic flux ratio (in units of the critical value for collapse $\mu_{cr} \approx (4~\pi^{2}~G)^{-1/2}$, Nakano \& Nakamura (1978)\footnote{As in our previous work, we use the value derived by Nakano \& Nakamura (1978) for $\mu_{cr}$, which is derived from a linear stability analysis and which is potentially more adapted to our initially uniform density configuration threaded by a unidirectional magnetic field. The value of $\mu_{cr}=1/(63~G)^{1/2}$ derived by (Mouschovias \& Spitzer 1976) from non-linear calculations differs from the first quantity by a factor of $25$ per cent.} and the beta plasma values are $B_{cl}=14.5, 4.6~\mu G$, $\mu_{cl}=2.8, 8.8$, and $\beta_{p}=0.1, 1$, for simulations B1, and B2 respectively. The Poisson equation is solved, right from the beginning of both simulations, using the conjugate gradient method implemented in RAMSES.  In this work, we restrict our analysis to magnetized cloud models since, aside from the now commonly accepted fact that clouds and cores are indeed magnetized (e.g., Crutcher et al. 2004),  we have shown in some of our previous work that cores formed in magnetized cloud provide the best match to the available observational constraints such as the exponent of virial parameter-mass relation of the cores (Dib et al. 2007a) and their lifetimes (Galv\'{a}n-Madrid et al. 2007) (particularly for the nearly magnetically-critical cloud).

In order to be able to handle the large data cubes when performing the clump finding, we have binned the simulations such as to generate $1024^{3}$ uniform data cubes for each of the time-steps of the simulations we have analyzed. Cores are identified on the uniform grids using a clump-finding algorithm based on a density threshold and a friend-of-friend criterion similar to the ones used in Dib et al. (2007,2008a,2008b), Dib \& Kim (2007), Audit \& Hennebelle (2005), and Hennebelle \& Audit (2007). We have performed a clump finding directly in the three-dimensional space using a variety of density thresholds ranging from $10~n_{aver}$ to $1280~n_{aver}$ with increments by a factor of 2. We make the simple assumption that a threshold of $20~n_{aver}=10^{4}$ cm$^{-3}$ is of the order of the excitation density for the NH$_{3}$ ($J-K$)=(1,1) transition, whereas a threshold of $160~n_{aver}=8 \times 10^{4}$ cm$^{-3}$ is of the order of the excitation density for the (1-0) N$_{2}$H$^{+}$ emission line. Below these values, it is assumed that the molecules are under-excited and that there is no emission in the considered lines. The clump-finding algorithm takes into account the periodicity of the simulation box and thus, of cores that extend across the box boundaries. 

\section{GENERAL PROPERTIES OF THE SIMULATIONS}\label{general}

\begin{figure}
\plotone{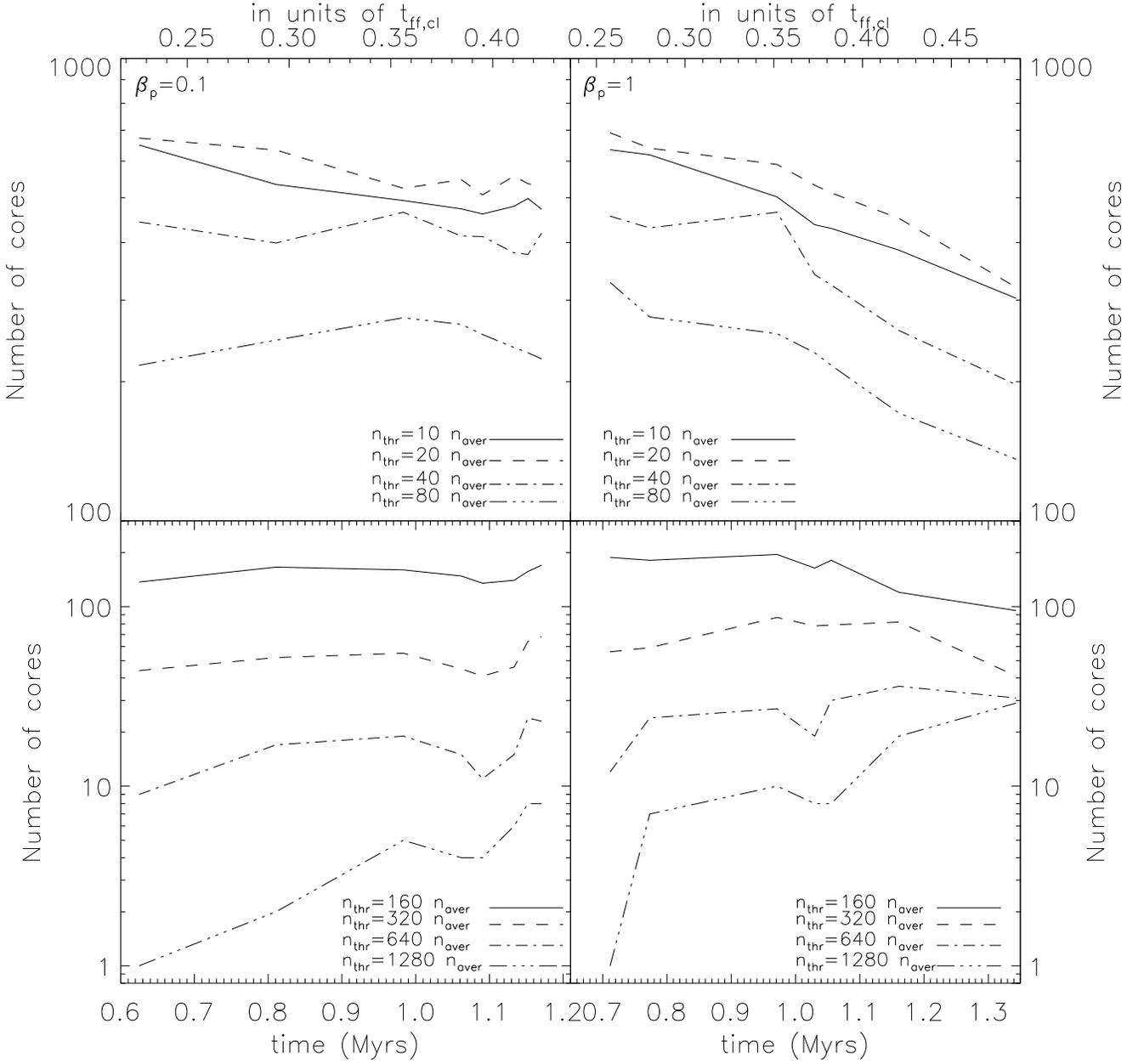} 
\caption{Time evolution of the number of cores identified at different density thresholds in the molcular cloud models B1(left, model with $\beta_{p}=0.1$) and  B2 (right, model with $\beta_{p}=1$).}
\label{fig1}
\end{figure}

\begin{figure}
\plotone{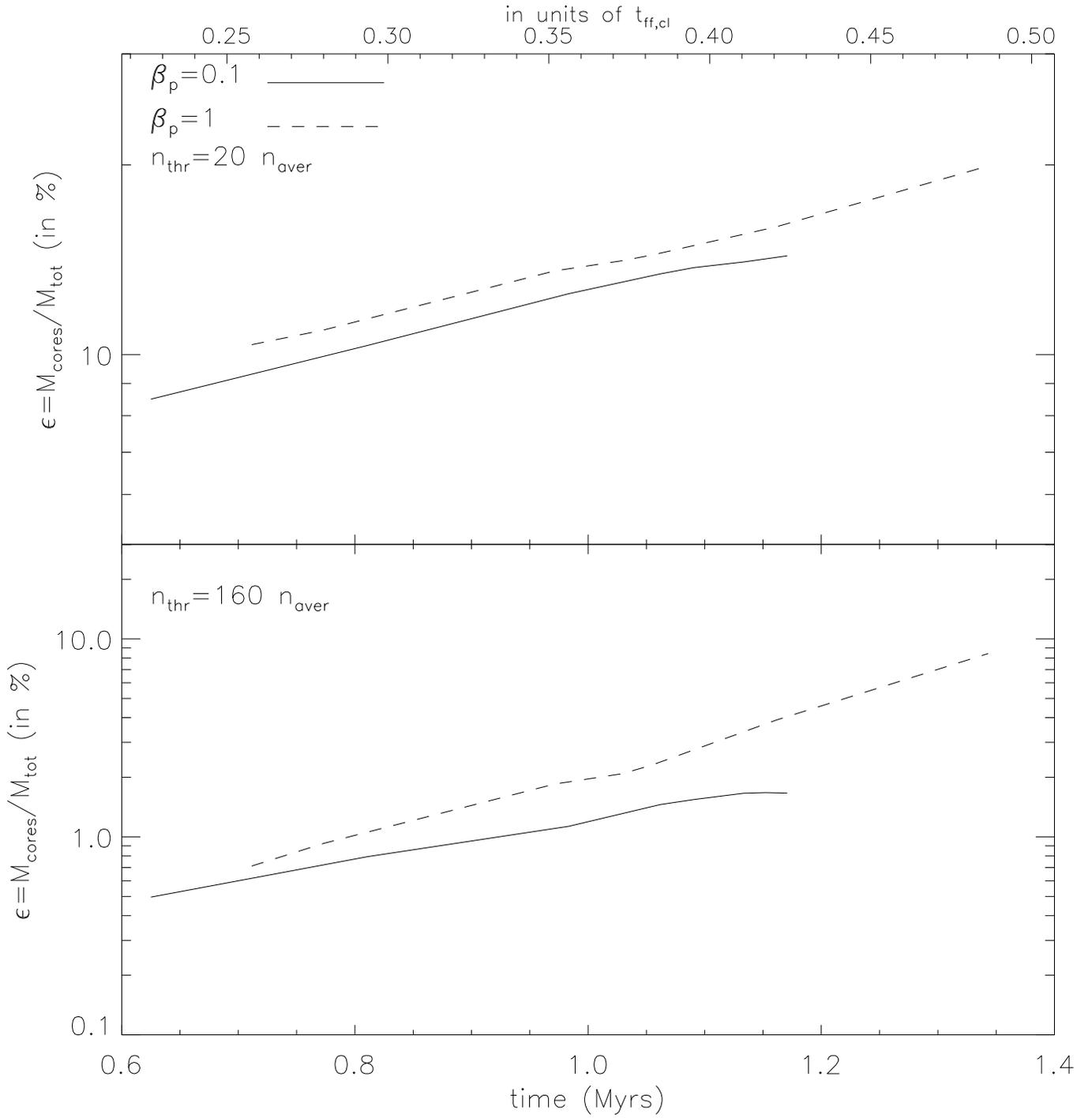}
\vspace{1.5cm}
\caption{Time evolution of the fraction of the simulation box mass that is found into dense cores identified at the thresholds of $n_{thr}=20~n_{aver}$ (C20 cores) and $160~n_{aver}$ (C160 cores) in the two simulations B1 and B2.}
\label{fig2}
\end{figure}

\begin{figure}
\plotone{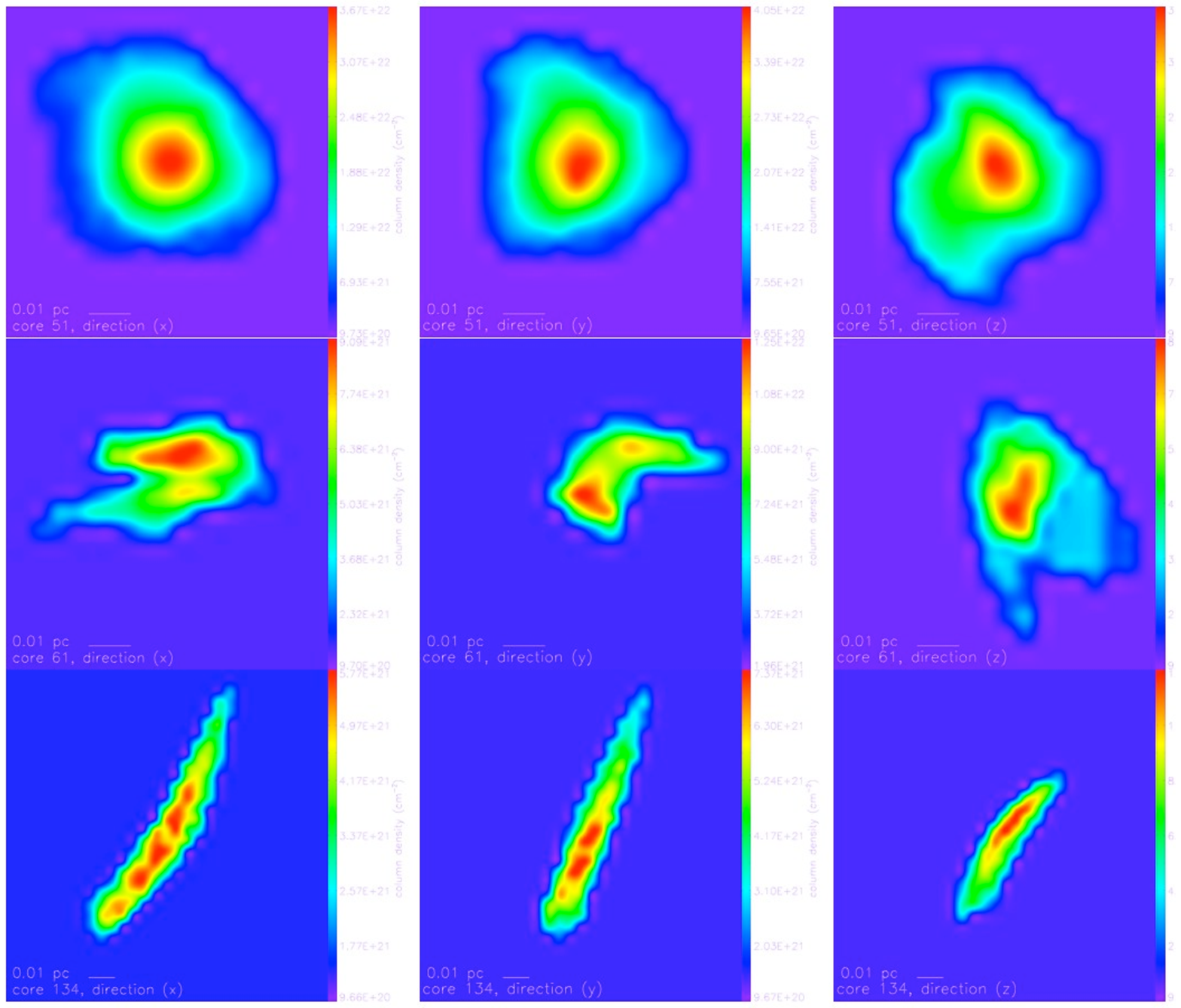}
\caption{Column density maps along the three directions of the simulation box for a few selected C160 cores at $t=1.152$ Myrs $=0.417~t_{ff,cl}$ in the simulation with $\beta_{p}=0.1$ (B1).}
\label{fig3}
\end{figure}

\begin{figure}
\plotone{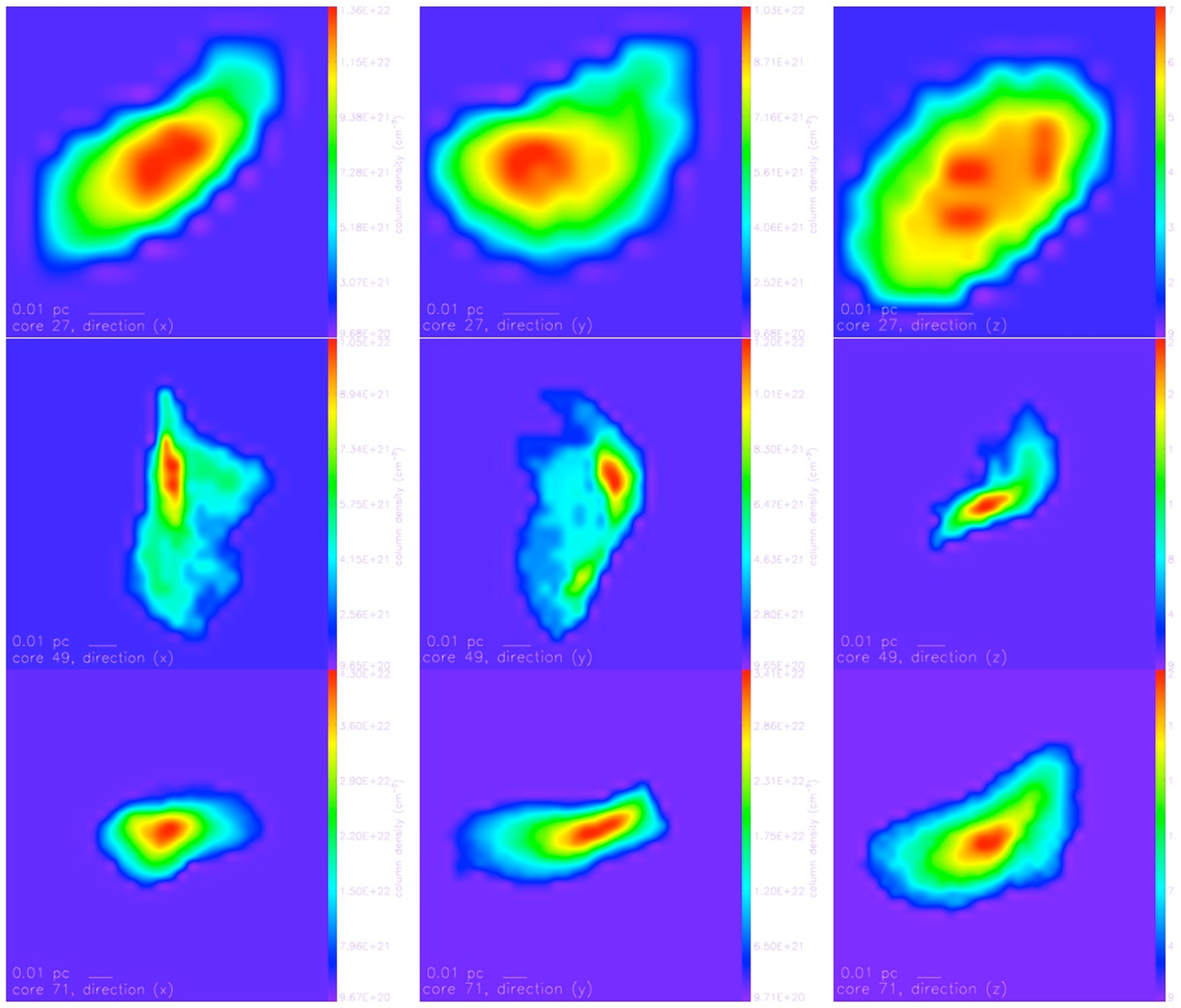}
\caption{Column density maps along the three directions of the simulation box for a few selected C160 cores at $t=1.159$ Myrs $=0.420~t_{ff,cl}$ in the simulation with $\beta_{p}=1$ (B2). }
\label{fig4}
\end{figure}

In this section, we briefly summarize some of the basic properties of simulations B1 and B2 in terms of their respective populations of cores identified at the selected density thresholds. As time evolves, turbulence decays in the box and the sonic Mach number decreases from ${\cal M}=10$ at $t=0$ to ${\cal M}$=[5.28, 3.54, 2.97, 3.08, 3.28, 3.36] at $t=[0.087, 0.226, 0.293, 0.356, 0.410, 0.424]~t_{ff,cl}$ is run B1 and to ${\cal M}$=[3.61, 3.37, 2.89, 2.89, 2.59, 2.73] at $t=[0.257, 0.280, 0.351, 0.373, 0.420, 0.486]$ in model B2. By $t \sim 0.42~t_{ff,cl}$, the $rms$ Mach number has decayed faster in model B2 to a value of $2.6$ and which is comparable to the $rms$ Mach number in the Ophiucus cloud (e.g., Enoch et al. 2007 based on the COMPLETE survey, Ridge et al. 2006) as compared to a value of $\sim 3$ in run B1 which is comparable to the $rms$ Mach number in the Perseus molcular cloud (Enoch et al. 2007, Pineda et al. 2008)\footnote{The velocity dispersions quoted by Enoch et al. (2007) in the Perseus and Ophiucus molecular clouds are averaged values measured from all the pointings in each cloud. This averaged velocity dispersion, and the associated Mach number calculated assuming a gas temperature of $10$ K can be associated with the average 'width' of the cloud along the line of sight which is expected to be of the order of a few pc }. In both runs, the Mach numbers increases again at later stages because the $rms$ velocity becomes increasingly dominated by the large velocity components in the collapsing cores. Tabs.~\ref{table1}-\ref{table4} display the $rms$ Mach numbers, the numbers of C20 cores (cores identified at a density threshold of $20~n_{aver}$) and the C160 cores (cores identified at a density threshold of $160~n_{aver}$), the peak density in the box in units of the average density, the median mass of the cores, their median sizes (taken to be the cubic root of the cores volumes), their median specific angular momentum, and their median rotational parameter (see below for definition) at two selected, nearly similar, epochs. At early epochs (i.e., when $t \lesssim~0.35~t_{ff,cl}$), the peak density in the simulation box, associated with the most massive cores is of the order of a few $10^{3}~n_{aver} \sim$ a few $10^{5}$ cm$^{-3}$ which is characteristic of the densities of gravitationally bound cores (e.g., Lee et al. 2001; Kandori et al. 2005). At $t \sim 0.42~t_{ff,cl}$, the most massive cores in the simulations are in a stage of free-fall collapse and the peak density is of the order of a few $10^{5} ~n_{aver} \sim$ a few $10^{7}$ cm$^{-3}$ which is similar to the densities of cores showing signature of infall (e.g., J{\o}rgensen et al. 2002). Fig.~\ref{fig1} shows the time evolution of the number of cores in the simulation box in models B1 and B2 at the thresholds of $n_{thr}=10, 20, 40, 80, 160, 320, 640,$ and $1280$ $n_{aver}$, respectively\footnote{The data starts in figures 1, 2, and 3 starts at slightly different points in time as we have dumped the output at equal number of the simulations time-step which varies in time and from simulation to simulation. The data points were chosen such as to cover the epoch at which dense structures form and up to the point where a runaway collapse starts to set in in one of the cores.}. Fig.~\ref{fig2} displays the time evolution of the mass fraction of the box enclosed in the C20 cores and  the C160 cores in the two simulations. Fig.~\ref{fig3} and Fig.~\ref{fig4} display the column density, along the three directions of the box, of a few selected C160 cores at $t=1.152$ Myrs $=0.417~t_{ff,cl}$ in simulation B1 and at $t=1.159$ Myrs $=0.420~t_{ff,cl}$ in simulation B2, respectively. The physical size of the cores is indicated by a marking of the $0.01$ pc scale at the lower left corner of each map. The cores show a variety of morphologies going from more roundish structures to more filamentary ones. Some of the cores exhibit a single density peak whereas others are already fragmented into two or several sub-cores. Several authors (e.g., Price \& Bate 2007, Hennebelle \& Teyssier 2008) have suggested that strong perturbations must be seeded in the clumps/cores in order to allow them to fragment in the presence of a magnetic field. Our simulations suggest that such strong perturbations which would allow for fragmentation, occur naturally in cores formed in a gravo-turbulent, magnetized, molecular cloud. 

There are a number of similarities between the two simulations and a number of noticeable differences. In both simulations, the number of cores identified at the lower density thresholds ($n_{thr}=10-80~n_{aver}$) declines as time advances, while the number of cores identified at the higher density threshold ($n_{thr}=160-1280~n_{aver}$) increases. The increase in the high density cores is simply due to the effect of gravity which generates more centrally peaked objects as time advances. On the other hand, the decrease in the low density cores, in addition to being a signature that some of them are not gravitationally bound and may disperse as earlier suggested by V\'{a}zquez-Semadeni et al. (2005a) and Dib et al. (2007a), is also an indication of core coalescence which depletes a fraction of these low density cores (Dib et al. 2007b, Dib 2007, Dib et al. 2008c). The number of cores in the stronger magnetized cloud simulation (B1) is observed to be initially smaller than in the strongly magnetically-critial cloud simulation B2 and the number of dense cores formed at the later stages in the evolution of the cloud is also observed to be smaller in the B1 cloud simulation than in the B2 model. This is also consistent with previous findings (V\'{a}zquez-Semadeni et al. 2005a and Dib et al. 2007a) which suggests that the number of cores formed in the cloud decreases when the cloud is more magnetically supported.

As can be observed in Fig.~\ref{fig2}, the mass fraction of the cores with respect to the mass enclosed in the box, $\epsilon=M_{cores}/M_{box}$ is larger and increases at a faster pace in the strongly supercritical simulation B2 than in the mildly supercritical one, B1. The definition of the core formation efficiency as being the mass enclosed in the cores per unit free-fall time of the cloud: 

\begin{equation}
CFE= \frac{\Delta \epsilon} {t_{ff,cl}}, 
\label{eq1}
\end{equation}

\noindent yields core formation efficiencies for the C20 and C160 cores, based on the variations of $\epsilon$ between the first and last time-step of each simulation in Fig.~\ref{fig2}, of $CFE$ (C20) $=0.29$ and $CFE$ (C160) $=0.059$, and $CFE$ (C20)$=0.41$ and $CFE$ (C160) $=0.33$ for simulations B1 and B2, respectively. The results for the strongly supercritical simulation (B2) are consistent with previous results (e.g., Li \& Nakamura 2004;  V\'azquez-Semadeni et al. 2005b, Nakamura \& Li 2005). Our results also show that the CFE is strongly reduced with an increasingly important magnetic field in the cloud, particularly for the case of the higher density cores.  It is useful to emphasize any comparisons between isothermal models such as ours and the observations remain valid as long as we are dealing with pre-stellar cores which are themselves embedded in a cloud that does not harbor strong sources of radiation (stars/protostars). The presence of radiation will unavoidably alter the structure of the cloud and reduce the core formation efficiency, and subsequently the star formation efficiency as recently pointed out by the simulations of Bate (2009) and Offner et al. (2009).

\section{ROTATIONAL PROPERTIES OF MAGNETIZED CORES}\label{comp}

After having defined the cells that belong to the same core, we calculate the components of the angular momentum of the core, ${\bf J}_{c}$ around the center of mass using the standard expression: 

\begin{equation}
{\bf J}_{c}=\sum_{i}  {\bf r}_{i} \times m_{i}~{\bf v}_{i},
\label{eq2}
\end{equation}

where ${\bf r}$ is the distance from each cell to the center of mass, $m_{i}$ the mass of the cell, and ${\bf v}_{i}$ is the relative velocity of the cell with respect to the core's center of mass. The specific angular momentum is then calculated as being:
 
\begin{equation}
 j_{3D}= \frac{{\bf |J}_{c}|} {M_{c}},
\label{eq3}
\end{equation}
 
where $M_{c}$ is the mass of the core. The rotational parameter $\beta_{rot}$, calculated in the three-dimensional space, is usually defined as being: 
 
\begin{equation}
\beta_{rot}= \frac{E_{rot}}{|E_{grav}|}, 
\label{eq4}
\end{equation}
 
where $E_{grav}$ and $E_{rot}$ are the core's gravitational energy and kinetic energy stored in rotational motions, respectively. The latter is given by its basic definition: 

\begin{equation}
E_{rot} = \int_{V_{c}} \frac {1}{2}  \rho {\bf r}^{2}  {\bf \omega}^{2} dV,
\label{eq5}
\end{equation}

where $V_{c}$ is the volume of the core, and ${\bf \omega}$ is the local angular velocity given by:

\begin{equation}
{\bf  \omega} = \frac { {\bf r}\times {\bf v}} {|{\bf r}|^{2}}.
\label{eq6}
\end{equation}

The gravitational energy $E_{grav}$ of the cores is observationally approximated by  $E_{grav}=-q G M_{c}^{2}/R_{c}$, where $M_{c}$ is the mass of the core and $q$ is a positive number which accounts for the core morphology and inner mass distribution. However, as $\beta_{rot}$ is intended to represent a volume integrated estimate of the balance between centrifugal forces and effective gravitational forces taking into account that dense cores in molecular clouds are not isolated objects, and in order not to make any assumption on the values of $q$ in the 3D estimate, it is more adequate, from a theoretical point of view, to replace $E_{grav}$ by the quantity $W$ which is slightly different than $E_{grav}$ (e.g., Dib et al. 2007a) and which is given by:

\begin{equation}
W= - \int_{V_{c}} \rho {\bf r} \left( \frac {\partial \phi} {\partial r }  \right) {\bf e}_{r} dV,
 \label{eq7}
 \end{equation}
 
where $\phi$ is the gravitational potential. Note that $W$ is not exactly equal to the volume gravitational energy, $E_{grav}$, because the true gravitational potential is a result of the distribution of matter inside the core and outside of it (e.g., the parent cloud). However, as already shown in Dib et al. (2007a), the essential part of the gravitational acceleration in the core is due to the mass contained inside the core and in general $W \approx E_{grav}$. Thus, our definition of $\beta_{rot}$ is: 

\begin{equation}
\beta_{rot}= E_{rot}/ |W_{grav}|.
\label{eq8}
\end{equation} 

It is important to stress that in a time dependent model of a turbulent cloud with gravity, the statistical properties of  the cores formed in the cloud may vary with time (e.g., Jappsen \& Klessen 2004; Dib et al. 2007a). In addition to angular momentum loss that may be due to the cores secular evolution (i.e., magnetic braking), the angular momentum of a core might be modified if the core enters a soft gravitational encounter with another/other core(s) and/or a physical merger (e.g., Larson 2010). Following a collision, the remnant merger may inherit a reduced or enhanced specific angular momentum. In our simulations, Fig.~\ref{fig1} shows that the number of cores detected at the lowest thresholds declines as a function of time while the number of cores identified at the highest thresholds increases. As some of the cores evolve into a stage of gravitational contraction, their average density will increase and they will be detected at higher density thresholds. However, as most of the cores identified at the thresholds of $10-80~n_{aver}$ are gravitationally unbound, and have generally long dispersion timescales (i.e., V\'{a}zquez-Semadeni et al. 2005), it is likely that the rapid decline in their numbers is due to merger events. Thus, in a dynamically evolving cloud, it is important to keep in mind that the overall angular momentum of the dense cores that form in the cloud will be re-distributed among the cores and the inter-core medium.

\subsection{Distributions of intrinsic specific angular momentum and rotational parameter}\label{distrib}

\begin{figure}
\plottwo{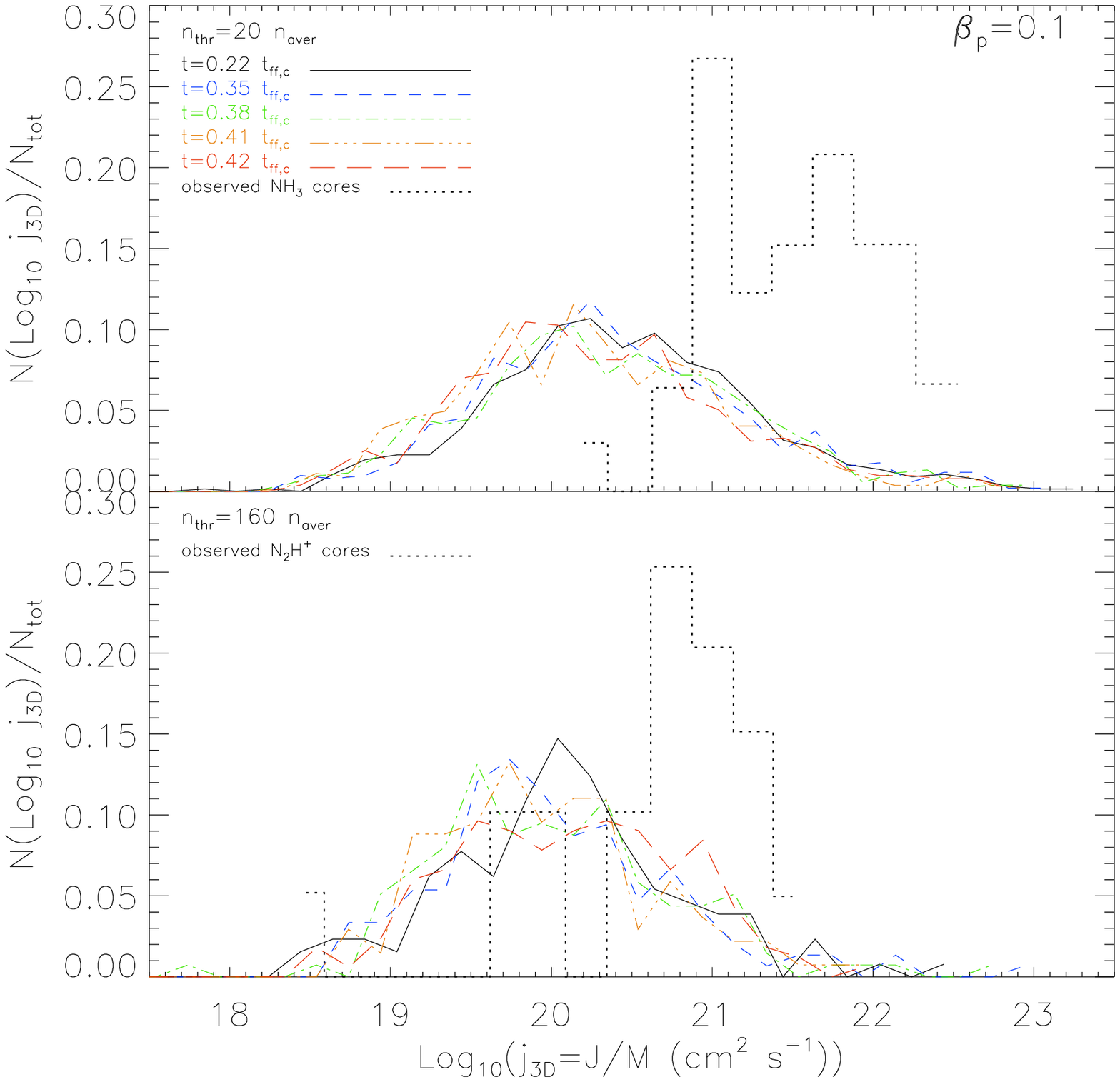} {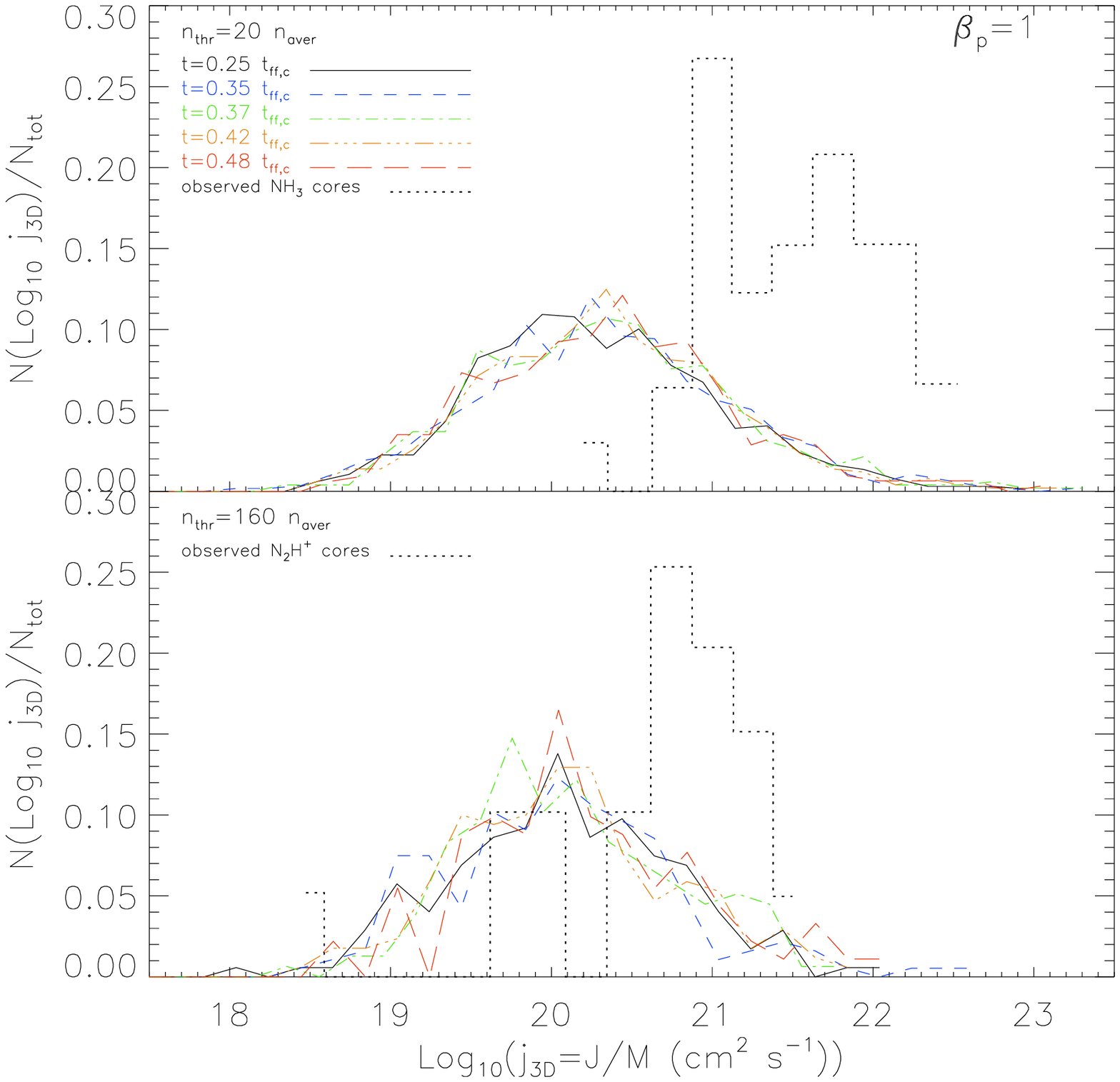}
\caption{Distributions of the specific angular momentum of the C20 and C160 cores at five different epochs in simulations B1 (left) and B2 (right). The NH$_{3}$ cores observations are those obtained by Goodman et al. (1993) and Barranco \& Goodman (1998), and the N$_{2}$H$^{+}$ ones are those obtained by Caselli et al. (2002a).}
\label{fig5}
\end{figure}

\begin{figure}
\plottwo{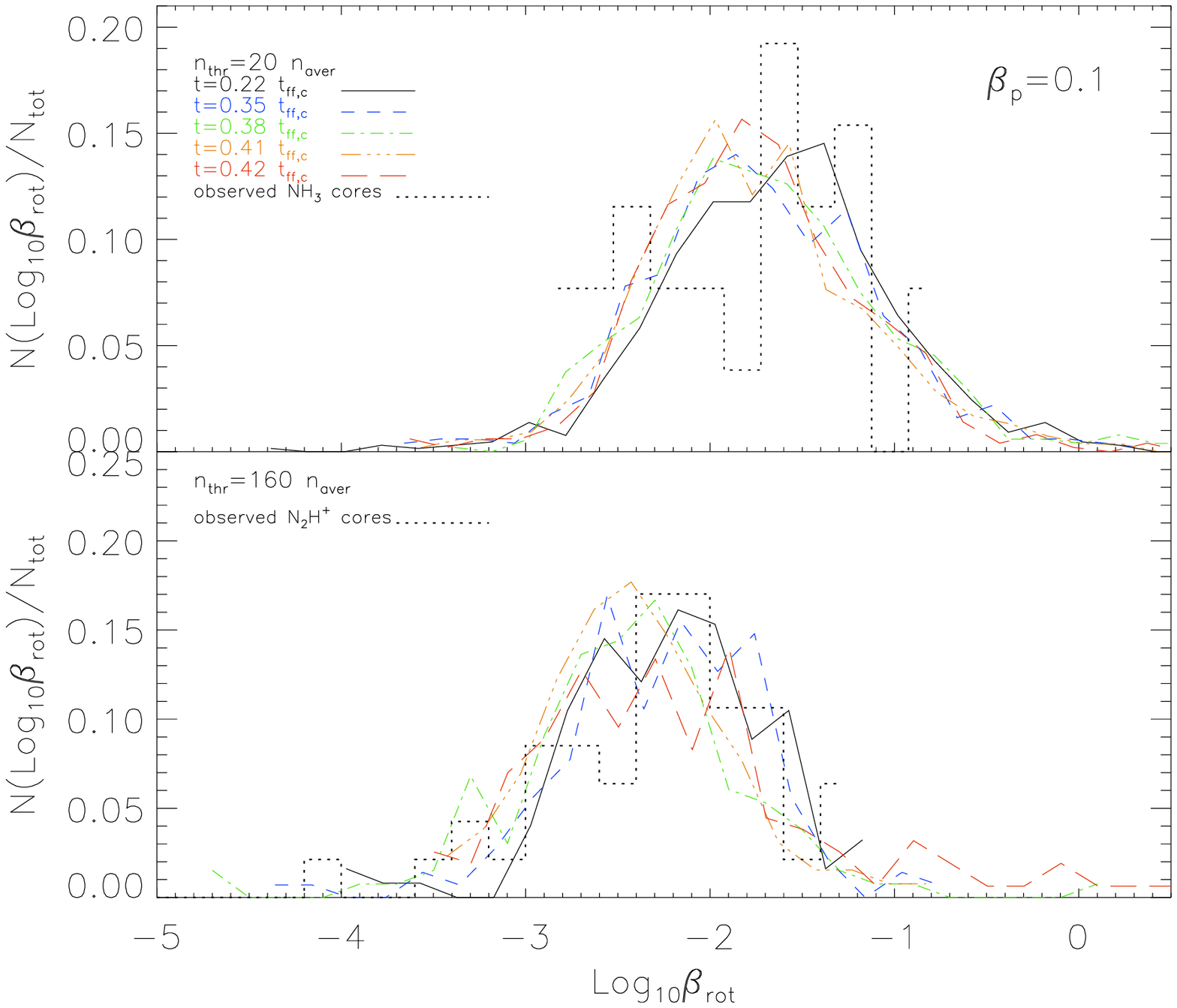}{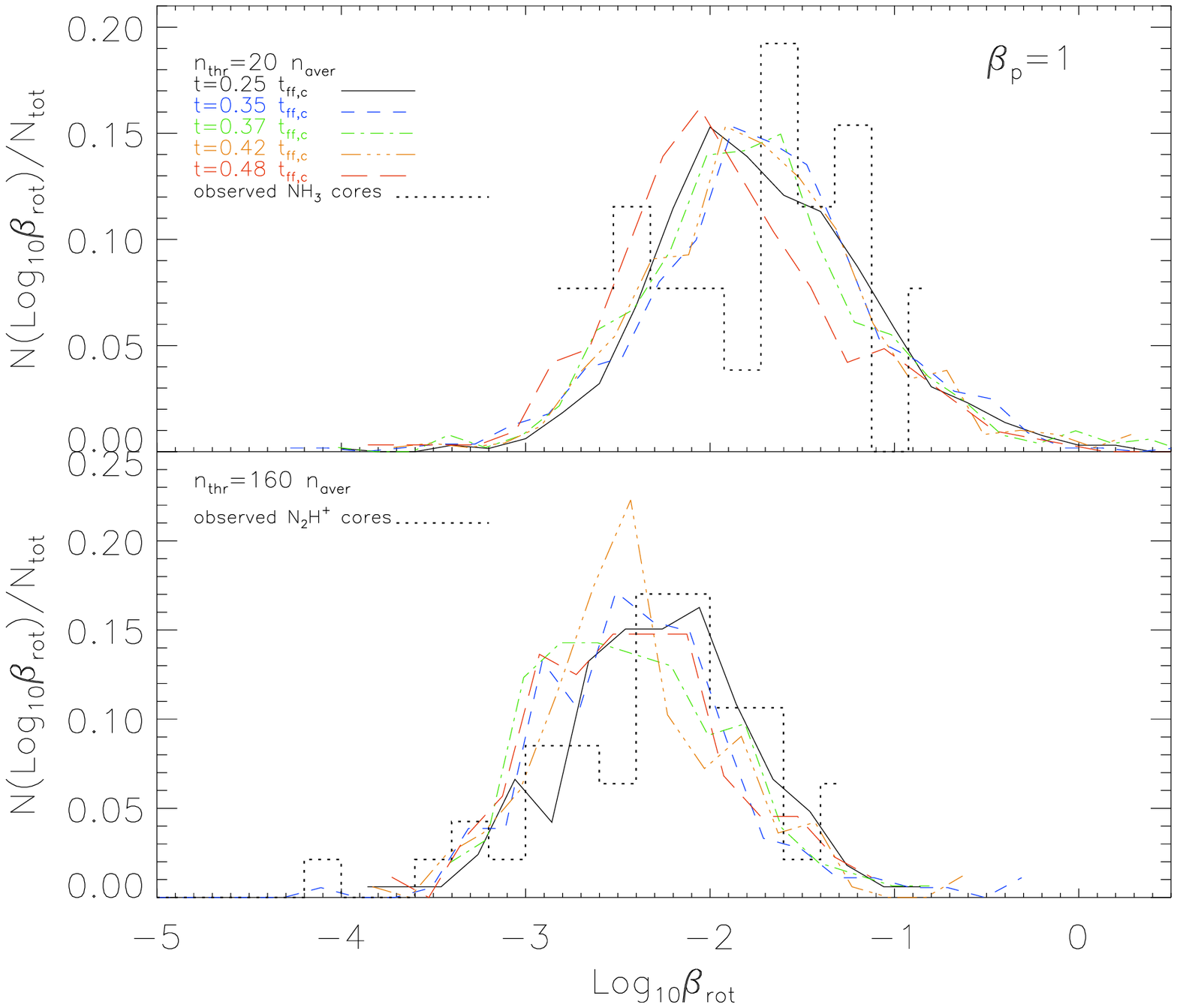}
\caption{Distributions of the $\beta_{rot}$ values of the C20 and C160 cores at five different epochs in simulation B1 (left) and B2 (right). The NH$_{3}$ cores observations are those obtained by Goodman et al. (1993) and Barranco \& Goodman (1998), and the N$_{2}$H$^{+}$ ones are those obtained by Caselli et al. (2002a).}
\label{fig6}
\end{figure}

\begin{figure}
\plotone{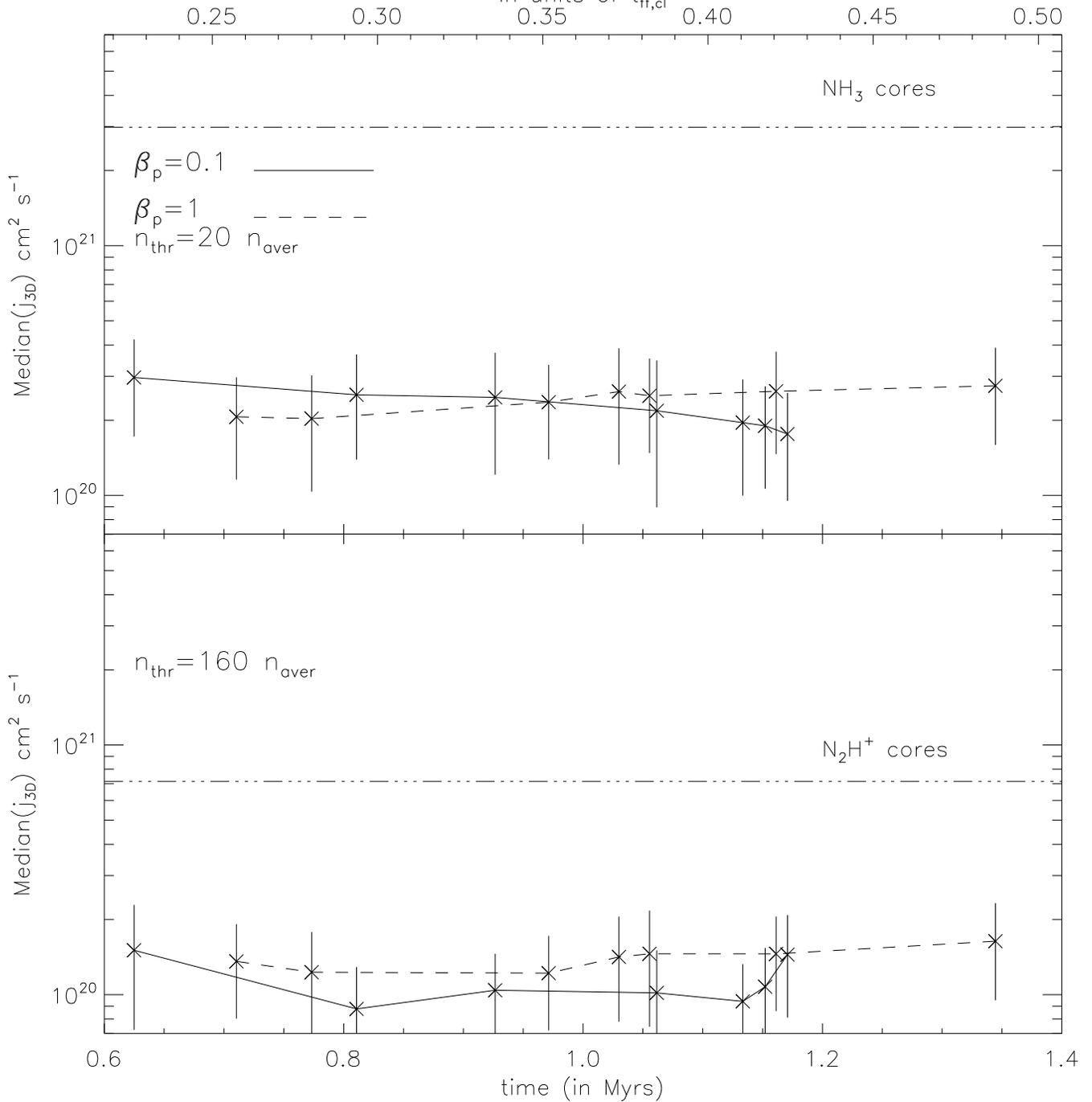}
\vspace{1cm}
\caption{Time evolution of the median value of the specific angular momentum for the C20 cores (top) and C160 cores (bottom) in simulations B1 and B2. In the top panel the median value of the specific angular momentum of the NH$_{3}$ of Goodman et al. (1993) and Barranco \& Goodman (1998) is over-plotted to the data. In the bottom panel the median value of the specific angular momentum of the N$_{2}$H$^{+}$ cores of Caselli et al. (2002a) is over-plotted to the data.}
\label{fig7}
\end{figure}

\begin{figure}
\plotone{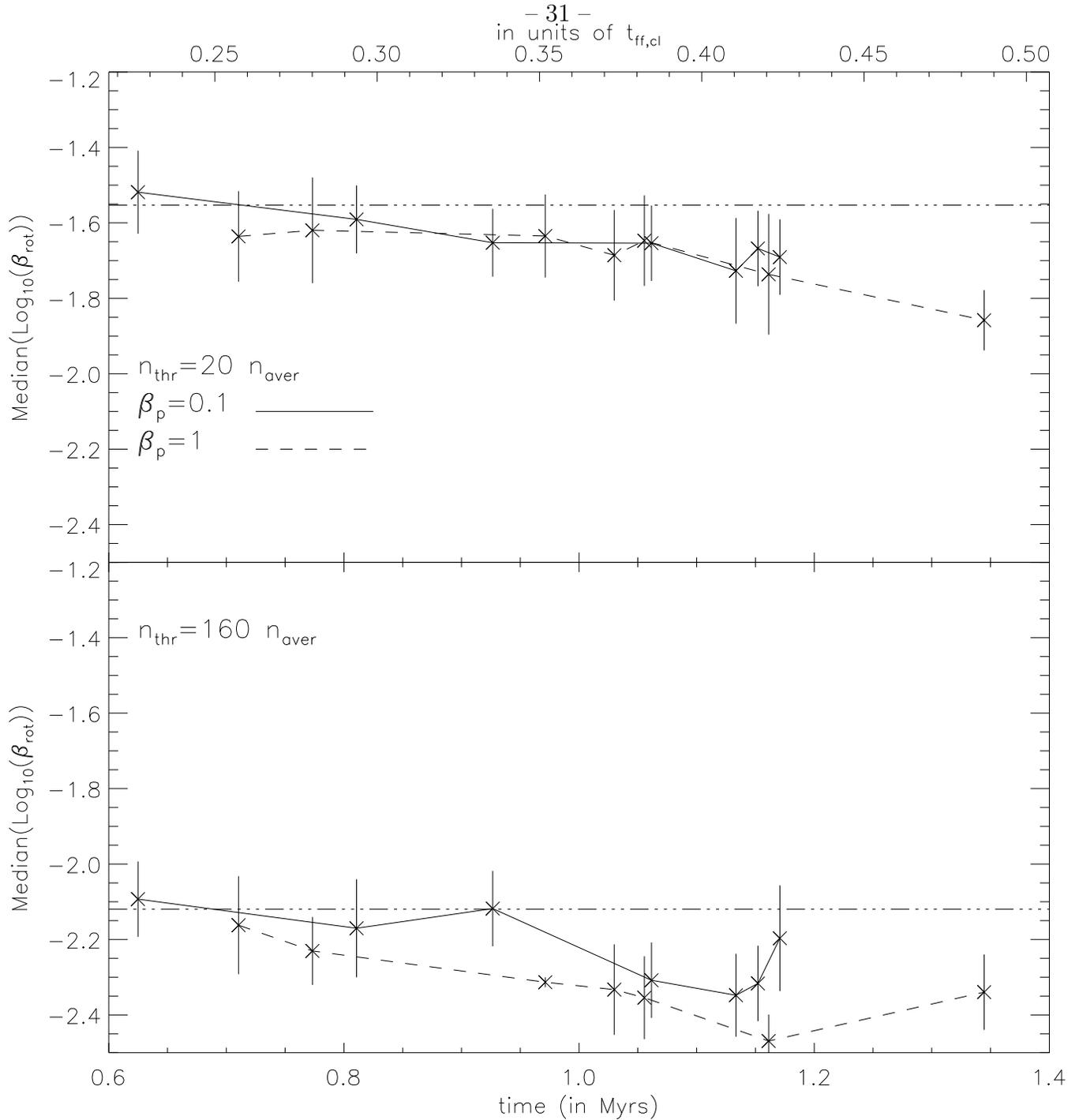}
\vspace{1cm}
\caption{Time evolution of the median value of the rotational parameter for the C20 (top) and C160 cores (bottom) in simulations B1 and B2. In the top panel the median value of the rotational parameter of the NH$_{3}$ cores of Goodman et al. (1993) and Barranco \& Goodman (1998) is over-plotted to the data. In the bottom panel the median value of the rotational parameter of the N$_{2}$H$^{+}$ cores of Caselli et al. (2002a) is over-plotted to the data.}
\label{fig8}
\end{figure}

Figs.~\ref{fig5} and \ref{fig6} display the time evolution of the normalized distributions of the intrinsic specific angular momentum $j_{3D}$ and of the rotational parameter $\beta_{rot}$ for the C20 and C160 cores identified in the two simulations. Albeit our models were not necessarily tailored to mimick any particular star forming region, it is temptative to over-plot to the numerical results the observationally derived NH$_{3}$ and N$_{2}$H$^{+}$ distributions of $j$ and $\beta_{rot}$ (dashed-line histograms). The NH$_{3}$ cores are those observed by Goodman et al. (1993) and Barranco \& Goodman (1998), whereas the N$_{2}$H$^{+}$ cores are those observed by Caselli et al. (2002a)\footnote{It should be noted that the NH$_{3}$ and N$_{2}$H$^{+}$ cores of Goodman et al. (1993) and Caselli et al. (2002a) and which are drawn from the sample of cores built by Benson \& Myers (1989) contain a mix of cores from different star forming regions including Ophiucus and Taurus that are located at different distances and which possess different dynamical characteristic (e.g., different Mach numbers).}. The median value of $j_{3D}$ for the C20 and C160 cores fluctuates in time but remains close to $j_{3D,med}$ (C20)=$2-3 \times 10^{20}$ cm$^{2}$ s$^{-1}$ and $j_{3D,med}$ (C160)=$8 \times 10^{19}-1.5 \times 10^{20}$ cm$^{2}$ s$^-{1}$ in simulation B1 and $j_{3D,med}$ (C20)=$2-3 \times 10^{20}$ cm$^{2}$ s$^{-1}$ and $j_{3D,med}$ (C160)=$1.5-2 \times 10^{20}$ cm$^{2}$ s$^{-1}$ in simulation B2. Similarly the median value of $\beta_{rot}$ for the two populations of cores fluctuates in time in the range of $\beta_{rot,med}$ (C20)=$(1.8-3) \times 10^{-2}$ and $\beta_{rot,med}$ (C160)=$(0.48-0.8) \times 10^{-2}$ in simulation B1 and $\beta_{rot,med}$ (C20)=$(1.4-2.3) \times 10^{-2}$ and $\beta_{rot,med}$ (C160)=$(0.34-0.68) \times 10^{-2}$ in simulation B2. Fig.~\ref{fig7} displays the time evolution of the median value of the C20 and C160 cores $j_{3D}$'s in simulations B1 and B2, and Fig.~\ref{fig8} displays the time evolution of the median values of $\beta_{rot}$ for the two populations of cores in the two simulations. The typical median absolute deviations of $j_{3D}$, shown as error bars in Fig.~\ref{fig7}, are of the order of 40-50 percent  and reflect the large width of the distributions in Figs.~\ref{fig5}. The median absolute deviations of $\beta_{rot}$, shown as error bars in Fig.~\ref{fig8}, are of the order of 30-40 percent. The median values of the specific angular momentum and of the rotational parameter for the C20 and C160 cores in the two simulations B1 and B2  are compared to those of the observed NH$_{3}$ cores and N$_{2}$H$^{+}$ cores in Fig.~\ref{fig7} and Fig.~\ref{fig8}, respectively. Albeit there is a certain overlap in the median values of $j_{3D}$ between simulations B1 and B2, the fact that the cores (particularly the C160 cores) in the more magnetized model have smaller values of $j_{3D}$ is consistent with the fact that magnetic braking leading to a loss of angular momentum is playing a more important role in the more strongly magnetized cloud model (B1) than in the less strongly magnetized cloud model (B2). We will address the issue of the origin of angular momentum loss by evaluating the effects of magnetic braking and the effects of gas accretion and coalescence of cores in a separate, dedicated, paper. 

A striking point in the comparison between the numerical specific angular momentum distributions of the C20 and C160 cores and the observed NH$_{3}$ and N$_{2}$H$^{+}$ distributions (i.e., Fig.~\ref{fig5} and Fig.~\ref{fig7}), respectively, is that the observations appear to overestimate the amount of the specific angular momentum by a factor of $\lesssim 10$. A similar trend is observed in the $\beta_{rot}$ distributions in Fig.~\ref{fig6} and Fig.\ref{fig8} albeit the shift in this case appears to be of the order of a factor $\sim 2$. Since, the observational determination of $\beta_{rot}$ relies on the determination of the specific angular momentum of the cores but also on the determination of the gravitational energy, many uncertainties can affect the determinations of $\beta_{rot}$ such as the assumed idealized morphologies of the cores and the idealized distribution of matter within them. In the following, we will therefore focus our attention on the discrepancy between the observed and numerically measured specific angular momentum distributions. Offner et al. (2008) calculated the values of the specific angular momentum for some of the cores in their simulations (at the resolution of 512$^{3}$) both in 3D and from projected velocity maps of the cores. They found that the 2D cores have specific angular momenta values that are systematically larger by a factor of $\sim 10$ than the values of $j$ for the 3D cores. Several effects may induce differences between a numerically calculated distribution of specific angular momentum and an observed one. In the present simulations, we used an initial $\it {rms}$ Ma number of $10$ and which decays to ${\cal M} \sim 3$ by $t \sim 0.42~t_{ff,cl}$ whereas the observed cores by Goodman et al. (1993) and Caselli et al (2002a) are sampled from different  molecular cloud complexes in which the dynamical conditions (e.g., the Mach number) are different. The most simple situation is that the parameters governing the dynamical evolution of  the molecular cloud in the models do not exactly match the dynamical conditions in the regions where the observed cores are formed and thus cores in the observations may have intrinsically more specific angular momentum than the simulated ones. The second effect may be related to the fact that the simulated cores are identified in intrinsic space whereas the observed cores are identified in position-position-velocity cubes. Although some differences in the cores populations may arise when using the two methods, most of the most massive cores will be detected in the PPV method with their masses modified by a correction factor of order unity (e.g., Smith et al. 2008). It is therefore unlikely that a factor of $\sim 10$ difference can originate from variations from the clump finding algorithm. A third possibility is that the totally different ways of calculating the specific angular momentum in the observations and in the simulations may stand behind this discrepancy. In order to test the latter hypothesis, it is necessary to generate synthetic velocity maps of the cores using individual projections, such as to eliminate any potential effect of blending of the cores along the line of sight, and measure the specific angular momentum from the projected velocity maps following the standard observational procedure. Unlike previous simulations, our models which have a high spatial resolution, and a large number of cores allow us to compare the entire distributions of the intrinsically measured specific angular momentum distributions to the ones derived from synthetic observations of the cores. 

\begin{figure}
\plotone{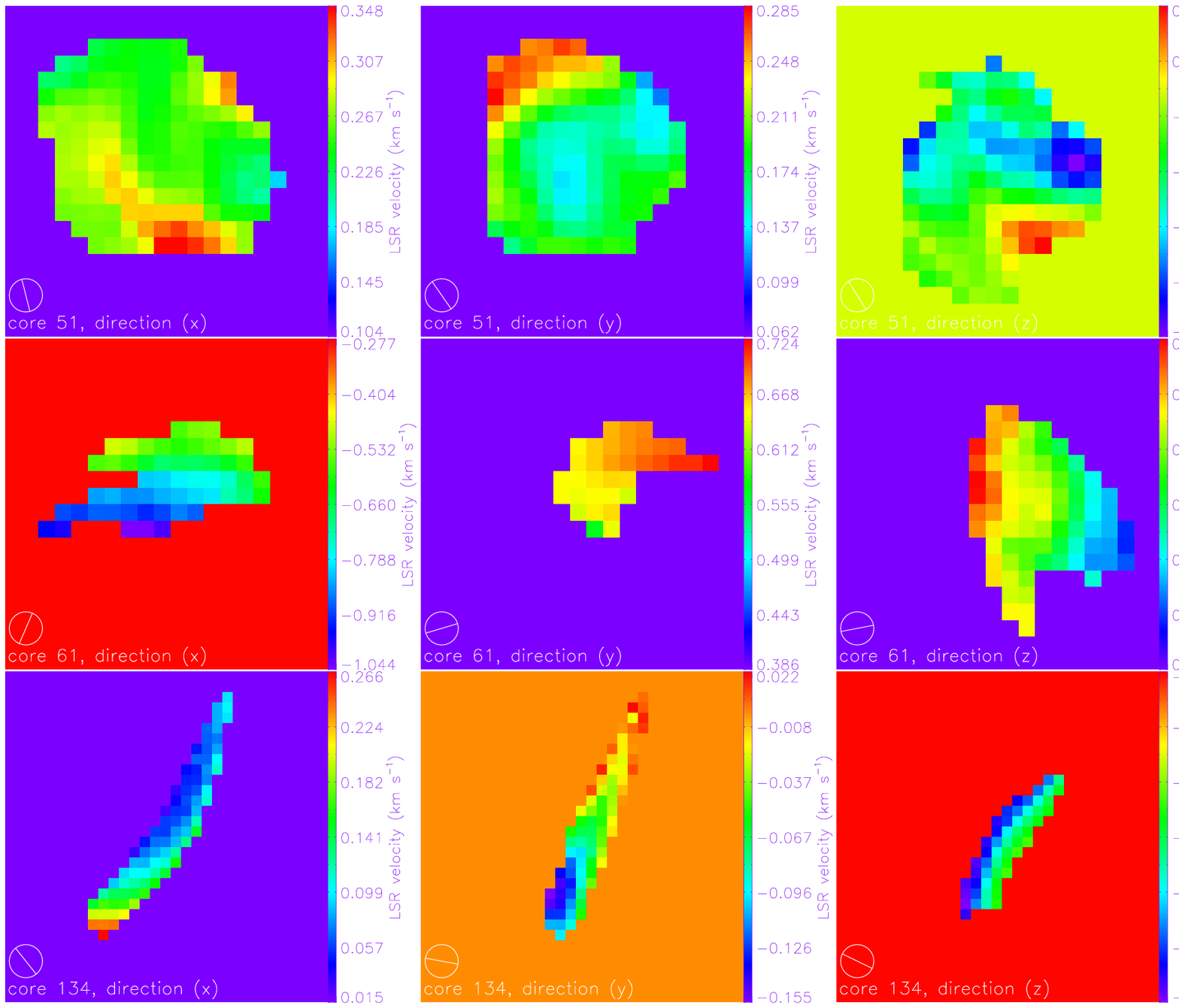}
\caption{Local standard of rest velocity maps (LSR) along the three directions of the simulation box for the same C160 cores shown in Fig.~\ref{fig4}. The line enclosed in the circle in the lowest left corner indicates the direction of the velocity gradient as obtained by VFIT.}
\label{fig9}
\end{figure}

\begin{figure}
\plotone{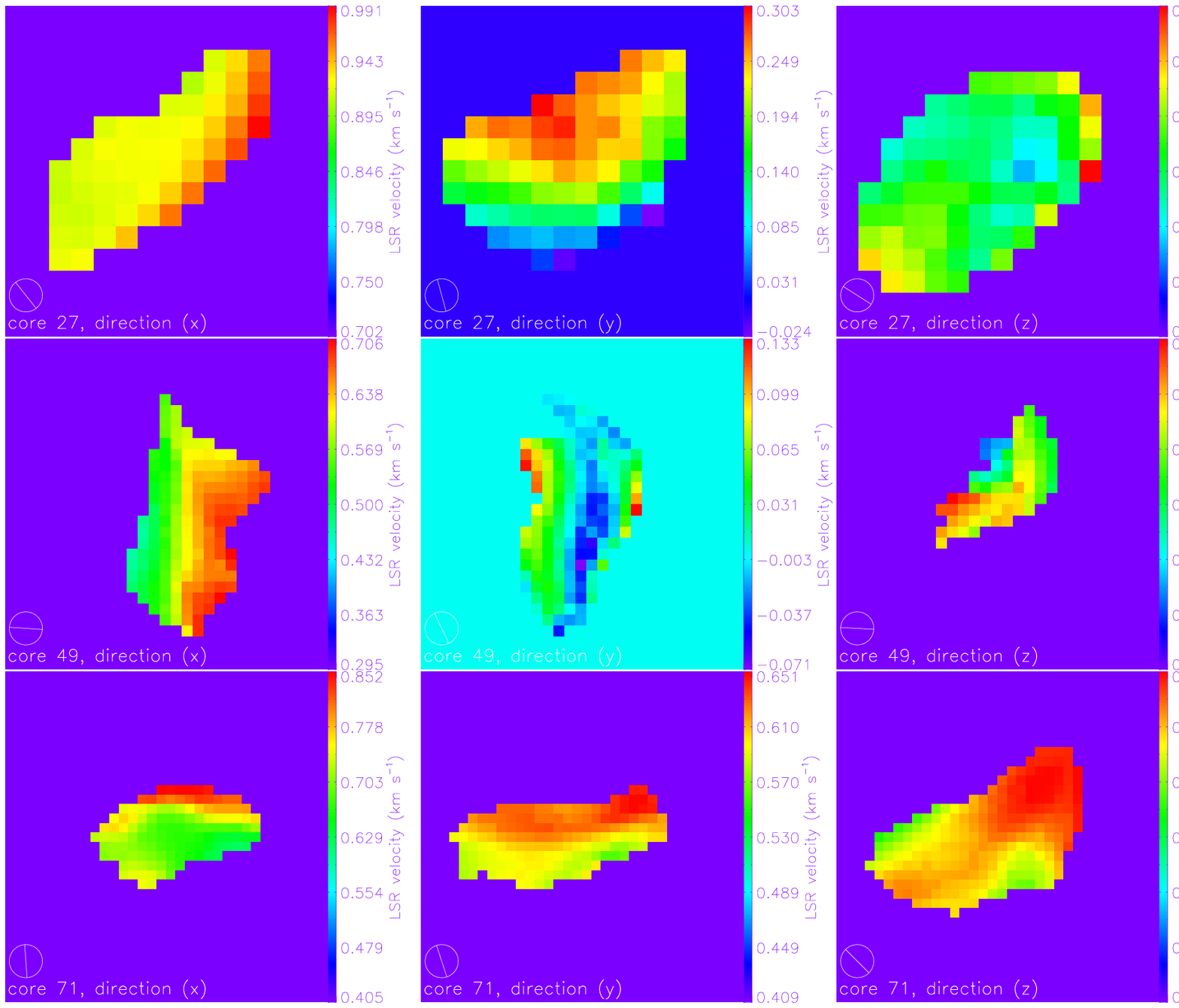}
\caption{Local standard of rest velocity maps (LSR) along the three directions of the simulation box for the same C160 cores shown in Fig.~\ref{fig5}. The line enclosed in the circle in the lowest left corner indicates the direction of the velocity gradient as obtained by VFIT.}
\label{fig10}
\end{figure}

We generate synthetic velocity maps of the C160 cores in both simulations at a nearly similar epoch (i.e., at $t=1.152$ Myrs $=0.417~t_{ff,cl}$ for simulation B1 and $t=1.159$ Myrs $=0.420~t_{ff,cl}$ for simulation B2) along the three main directions of the box. As stated above, the cores are projected individually such as to eliminate any effect of blending along the line of sight. The velocity in each pixel of the cores is the mean velocity in the line of sight. The velocity maps corresponding to the cores shown in Fig.~\ref{fig3} and Fig.~\ref{fig4} are displayed in Fig.~\ref{fig9} and Fig.~\ref{fig10}, respectively. The velocity maps exhibit a variety of features ranging from well ordered motions that can be assimilated to rotation, both for roundish and filamentary cores, to more complex dynamical configurations in which there is no clearly verifiable velocity gradient. In general, the velocity maps of the cores, especially when the same cores are seen along the three different projections, do not support the idea of simple rigid-body rotation. It is interesting to note that recent high spatial and spectral resolution observations using the Plateau de Bures interferometer by Csengeri et al. (2010) of five massive dense cores in Cygnus-X show a variety and level of complexity in their dynamical features similar to the ones observed in our synthetic velocity maps.     

In order to derive the velocity gradient from the synthetic velocity maps, we employ the program VFIT, developed by Goodman et al. (1993), which calculates the global velocity gradient in the map, $\cal G$ and its orientation. The uncertainty in the centroid velocity needed as an input by VFIT is taken to be the velocity dispersion of the velocity distribution along the line of sight. The direction of the velocity gradient for the velocity maps in Fig.~\ref{fig9} and Fig.~\ref{fig10} is shown by the line enclosed in the circle in the lowest left corner of each map. The projected specific angular momentum is given by $j_{2D}=J_{c,proj}/M_{c}$, where $J_{c,proj}$ and $M_{c}$ are the estimated projected angular momentum, and mass of the core, respectively. By making the assumption of uniform rotation, as is commonly done in the observations, one can write that $J_{c,2D}=I~\Omega$, where $I=p~M_{c}~R_{c,2D}^{2}$ is the moment of inertia of the core, $R_{c,2D}$ the value of the projected radius of the core taken to be the square root of its projected surface, and $p$ is a factor of order unity which depends on the morphology and radial density profile of the core. The angular velocity of the core is then estimated as being $\Omega = {\cal G}/{\rm sin} i$, where $i$ is the inclination of the rotation axis to the line of sight. Thus $j_{2D}$ is given by:

\begin{equation}
j_{2D}=p~R_{c,2D}^{2}~\frac{\cal G} {\rm sin~i}.
\label{eq9}
\end{equation}    

An assumption about the inclination of the cores and the value of ${\rm sin}~i$ has to be made. We follow Caselli et al. 2002 and use ${\rm sin}~i= \pi/ 4$, which is the average value of ${\rm sin}~i$ over all possible inclinations (Goodman et al. 1993 used ${\rm sin}i=1$. The result of this is just a systematic decrease of the estimated $j_{2D}$ by a factor of $4/\pi \approx 1.27$). In the observations, the assumption is also often made that cores have a uniform density which implies that $p=2/5$. Using these two approximations, $j_{2D}$ becomes:

\begin{equation}
j_{2D}=\frac{8}{5~\pi} R_{c,2D}^{2}~\cal{G}.
\label{eq10}
\end{equation}

\begin{figure}
\plotone{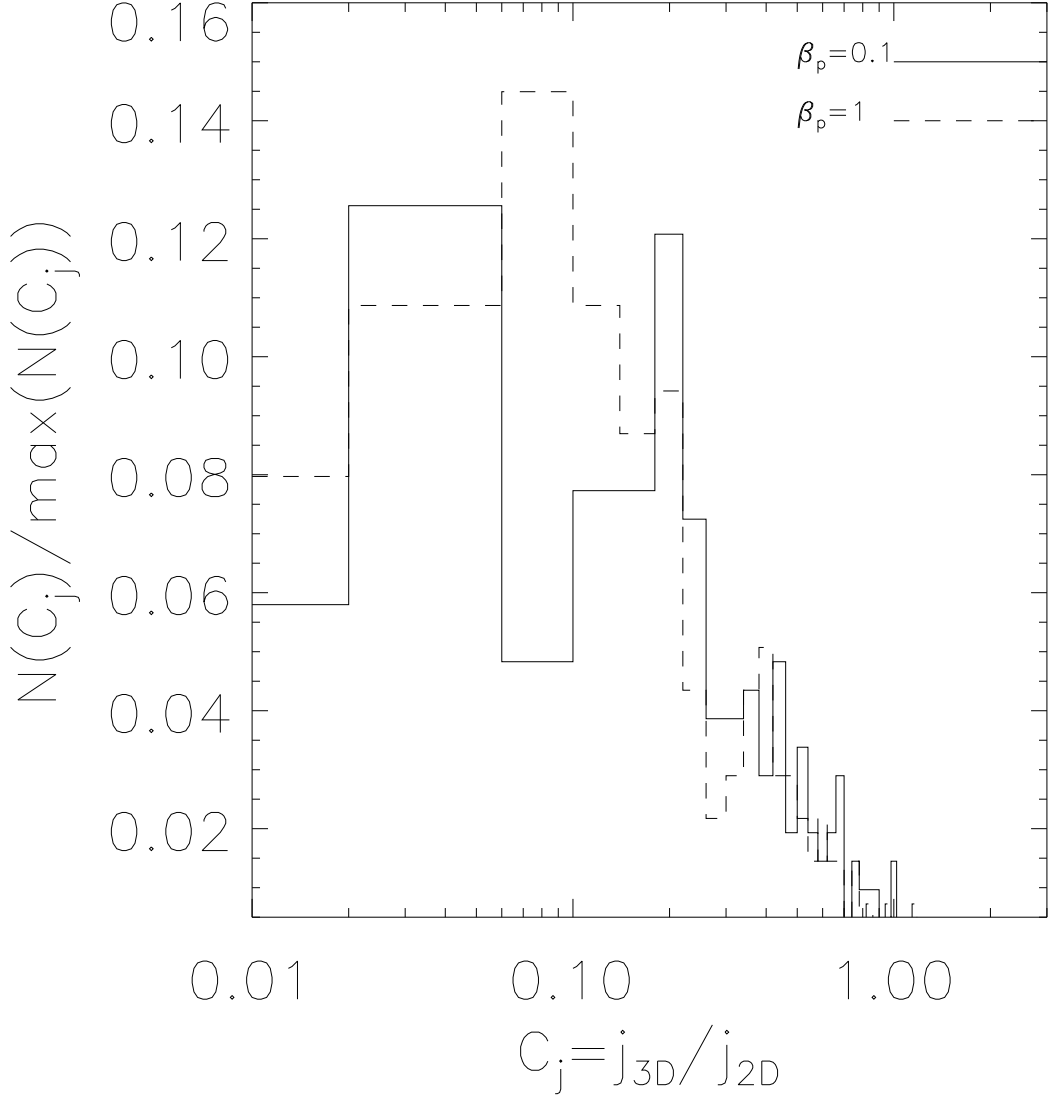}
\caption{Distribution of the ratio of the intrinsic specific angular momentum $j_{3D}$ to the specific angular momentum derived following the method adopted in the observations $j_{2D}$ for C160 cores identified at nearly similar epochs in simulations B1 (left) and B2 (right). The B1 data combines cores identified at $t=1.152$ Myrs $=0.417~t_{ff,cl}$ and $t=1.170$ Myrs $=0.424~t_{ff,cl}$ (in total, 69 cores and 207 velocity maps) and the B2 data displays the distribution for cores identified at $t=1.159$ Myrs $=0.420~t_{ff,cl}$ (in total, 46 cores and 138 velocity maps).}
\label{fig11}
\end{figure}

Fig.\ref{fig11} displays the ratio of the 3D and 2D estimates of the specific angular momentum $C_{j}=j_{3D}/j_{2D}$ for projections of the C160 cores along the three axis of the box at  $t=1.152$ Myrs $=0.417~t_{ff,cl}$ and $t=1.170$ Myrs $=0.424~t_{ff,cl}$ for simulation B1 (69 cores and 207 velocity maps in total for the two combined time-steps, left figure) and at $t=1.159$ Myrs $=0.420~t_{ff,cl}$ for simulation B2 (46 cores and 138 velocity maps in total, right figure). In order to avoid being affected by numerical noise, we have used C160 cores that have a minimum of 5 grid cells (on the $1024^{3}$ grid) in any given direction. From Fig.~\ref{fig11}, it is possible to observe that the value of $C_{j}$ is systematically $< 1$ for all cores and projections. The $C_{j}$ distribution peaks at $\sim 0.1$ in simulation B2 and displays an extended peak in the region $0.03-0.2$ in simulation B1. A two sided Kolmogorov-Smirnov test indicates that there is a probability of 38 percent for the two data sets in fig.~\ref{fig11} of being drawn from the same distribution, and 87 percent when the data is restricted to the range $C_{j}  \leq 1$. In terms of the median values, the median value of $C_{j}$ for the C160 cores in simulation B1 for the epoch displayed in Fig.~\ref{fig11} is $0.28 \pm 0.13$ and the median value of the C160 cores in simulation B2 is $0.22 \pm 0.11$. 

Whether the correction factor $C_{j}$ from $j_{2D}$ to $j_{3D}$ is considered in terms of its characteristic value ($\sim 0.1$) or in terms of its median value ($\sim 0.25$), the aim of Fig.~\ref{fig11} is to show that the observed shift between the observed and numerically derived distributions of the specific angular momentum as seen in Fig.~\ref{fig5} is most likely caused by the calculation method of the specific angular momentum using the global gradient method under the assumption of uniform rotation of the cores. We thus conclude, based on Fig.~\ref{fig5} and Fig.~\ref{fig11} that the observational determinations of the specific angular momentum tend to overestimate the true value of the specific angular momentum by a factor of at least $4-5$ but most likely produces an overestimate of $j$ by a factor of $8-10$. The origin of the 2D-3D discrepancy stems from the fact that the observational method is based on the global gradient method assuming uniform rotation, whereas the measurement of the angular momentum in the intrinsic space is a summation running over all parcels of gas in the cores with their more complex dynamical behavior.  

\subsection{The size-specific angular momentum relation }

A simplified method in theoretical works of assessing the specific angular momentum of cores is through the use of a size-specific angular momentum relation ($j_{3D} \propto R_{c}^{\lambda}$). If motions in the cores that are assimilated to rotational motions are purely due to turbulent motions following a Larson-like velocity dispersion-size relation of the form $\sigma_{c} \propto R_{c}^{\beta}$, then, the angular velocity $\Omega=\sigma_{c}/R_{c}$ will be given by $\Omega \propto R_{c}^{\beta-1}$ and the specific angular momentum by $j_{3D} \propto R_{c}^{1+\beta}$. For $\beta =0.38$ as observed initially by Larson (1981), the expected dependence of $j_{3D}$ is $\propto R_{c}^{1.38}$. On the other extreme hand, a rigid-body rotation (i.e., $\Omega$ is constant) implies that $j_{3D} \propto R_{c}^{2}$. 

\begin{figure}
\plotone{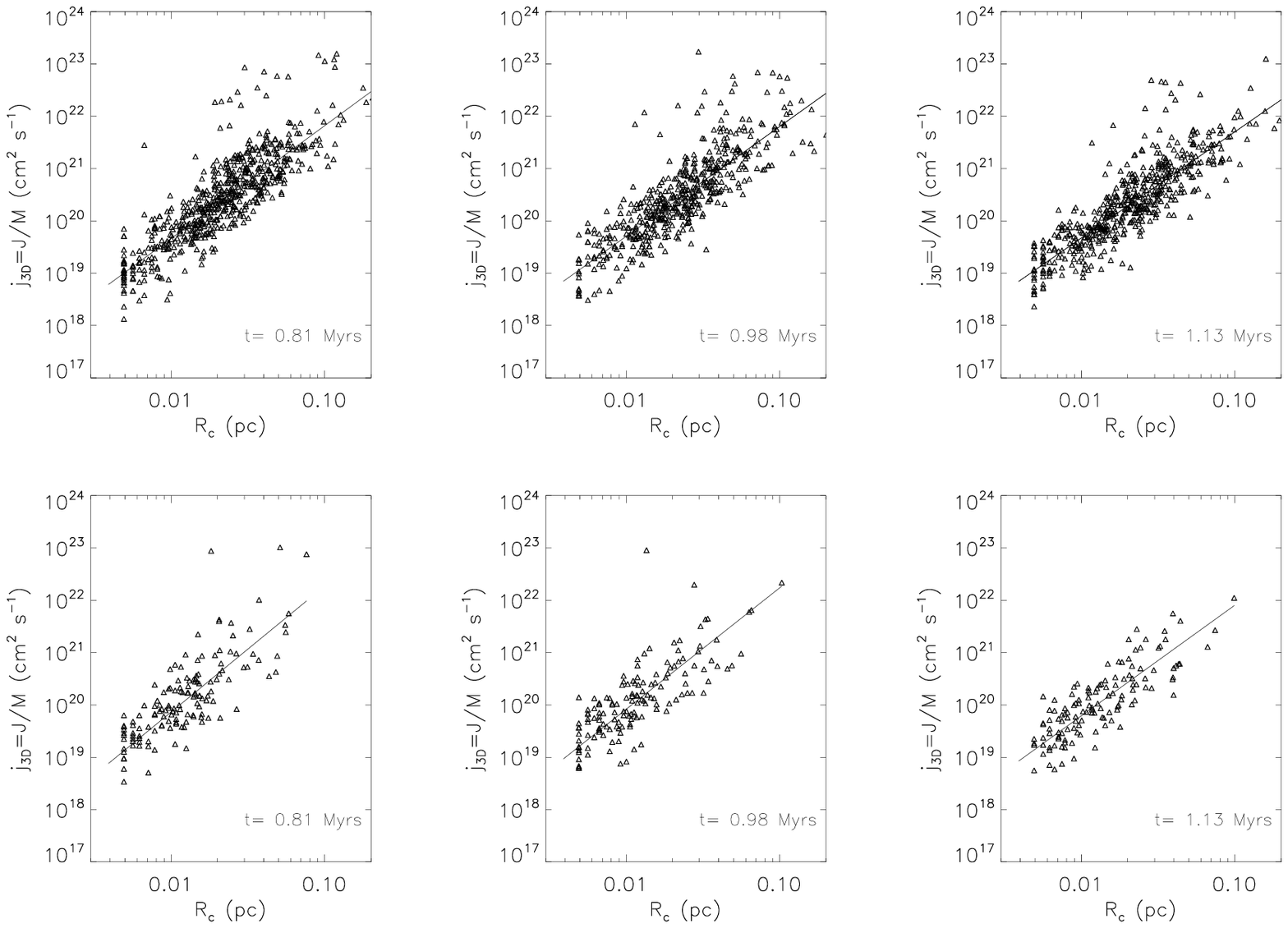}
\caption{Radius $R_{c}$-specific angular momentum $j_{3D}$ relation for the C20 cores  (upper row) and the C160 cores (lower row). identified in simulation B1 at a few selected epochs. $1~t_{ff,cl}=2.76$ Myrs.}
\label{fig12}
\end{figure}

\begin{figure}
\plotone{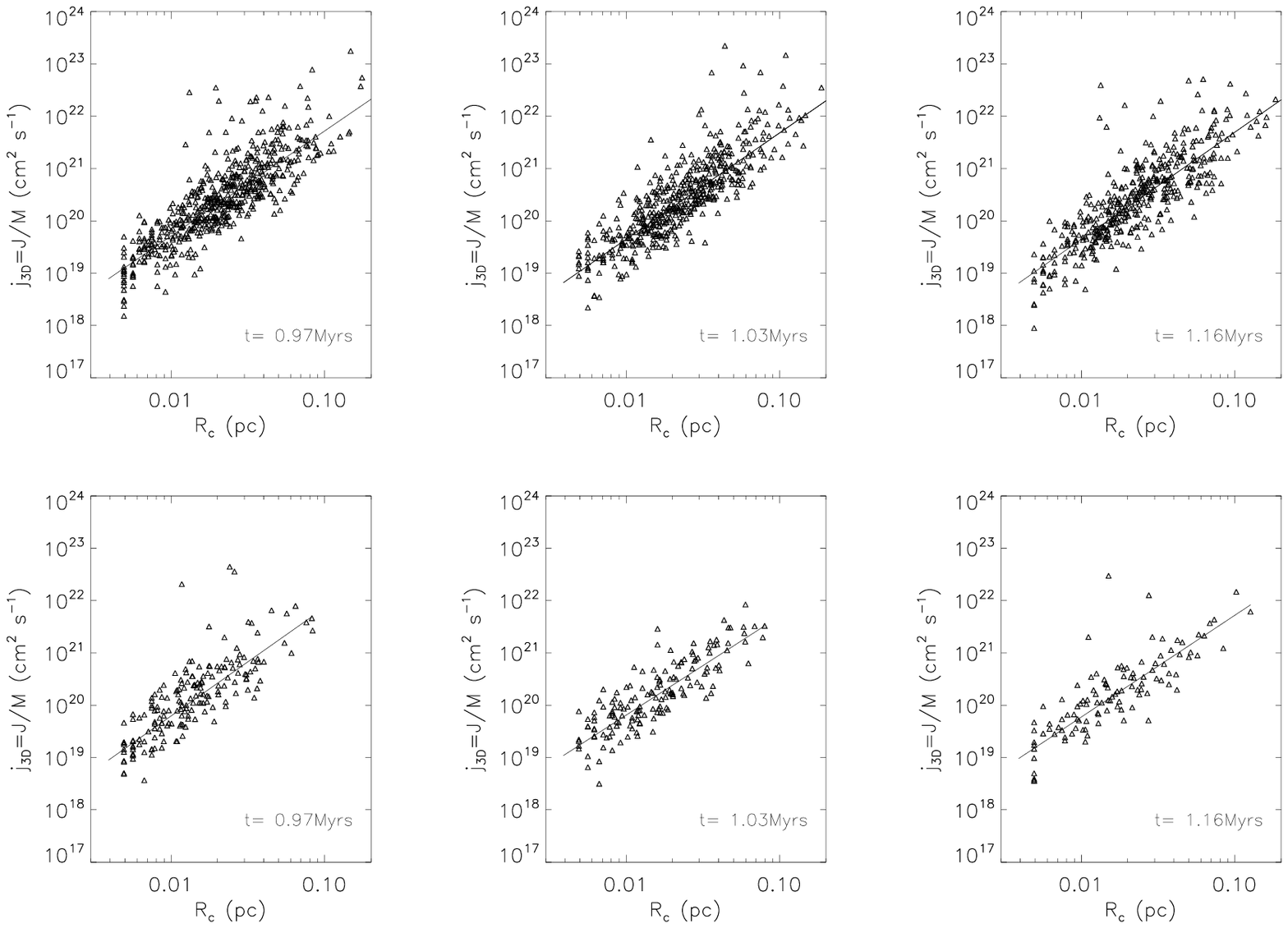}
\caption{Radius $R_{c}$-specific angular momentum $j_{3D}$ relation for the C20 cores (upper row) and  the C160 cores (lower row) identified in simulation B2 at a few selected epochs. $1~t_{ff,cl}=2.76$ Myrs. }
\label{fig13}
\end{figure}

\begin{figure}
\plotone{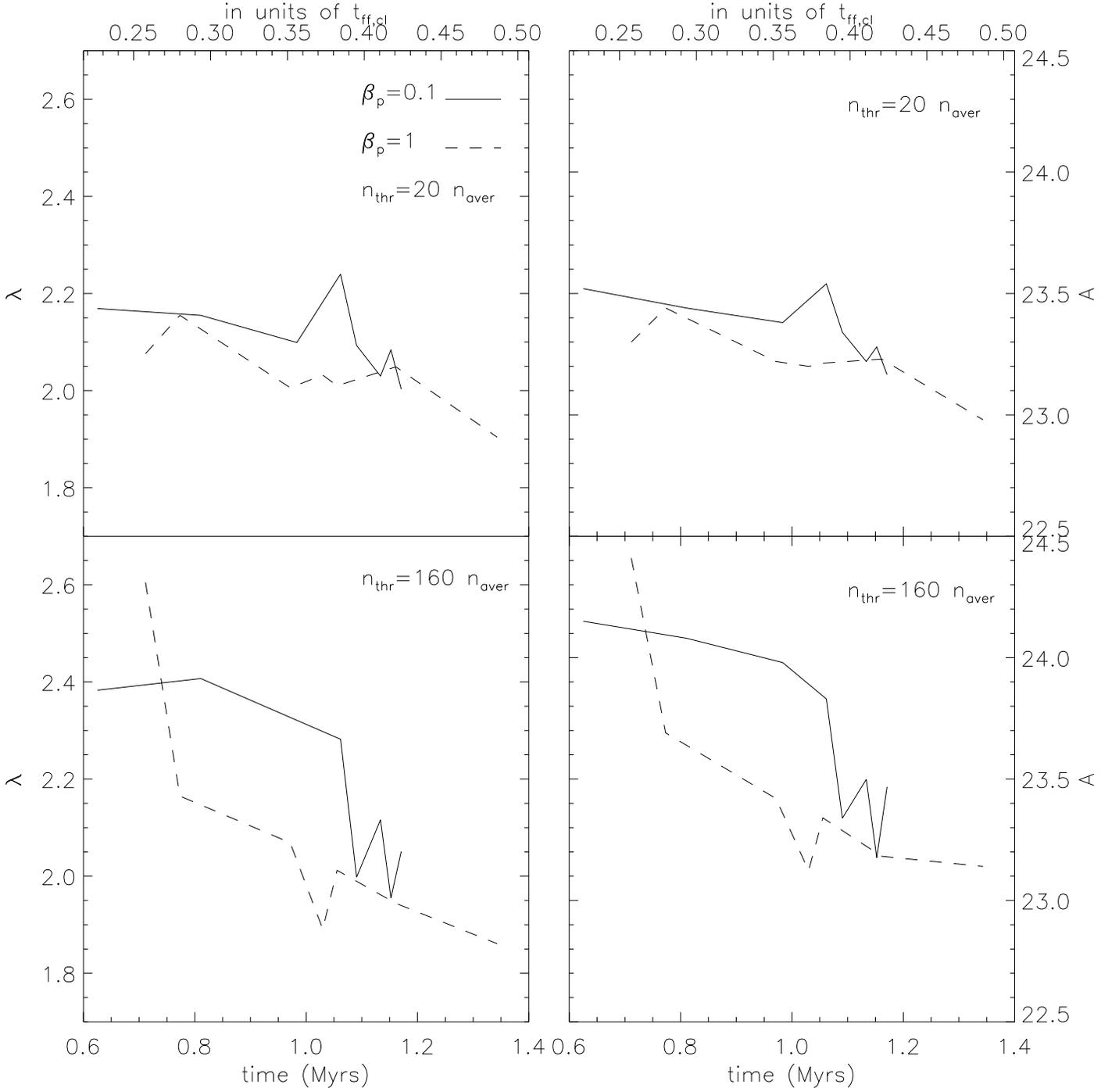}
\vspace{1cm}
\caption{Time evolution of the exponent and logarithm of the coefficient of the $Log(j_{3D})=A+ \lambda~Log(R_{c})$ relation for C20 and C160 cores identified in simulations B1 and B2. Time is given in physical units in the lower horizontal axis and in units of the free-fall time of the cloud in the upper horizontal axis.}
\label{fig14}
\end{figure}

Such relations have been derived observationally by several groups (e.g., Goldsmith \& Arquilla 1985; Goodman et al. 1993; Phillips 1999). Bodenheimer (1995) combined the data of Goldsmith \& Arquilla (1985) and Goodman et al. (1993) and found that the exponent of the size-specific angular momentum relation is of the order of $\lambda \sim 1.6$. Phillips (1999) compiled a large number of published molecular cloud data and their substructure of clumps and cores and found $\lambda$ to be of the order $\sim 1.43$ and close to $0.96$ for flattened systems. Fig.~\ref{fig12} and Fig.~\ref{fig13} display, at a few selected epochs, the intrinsic specific angular momenta $j_{3D}$ of the cores as a function of their characteristic size $R_{c}$ for C20 cores (top) and C160 cores (bottom) in simulations B1 and B2, respectively. We perform a robust least absolute deviation fit to the data displayed in Fig.~\ref{fig12} and Fig.~\ref{fig13} in addition to the same data at the additional epochs we analyze in order to derive the exponent and normalization of the relation $j_{3D}=A~R_{c}^{\lambda}$. Fig.~\ref{fig14} displays the time evolution of $\lambda$ and $A$ for the two populations of cores in the two simulations. Both $\lambda$ and $A$ are observed to decrease as a function of time and most importantly so for the C160 cores. The value of $\lambda$ is observed to decrease from around $\sim 2.5$ down to $\sim 1.8$ on a timescale of the order of $\sim 0.5 t_{ff,cl}$. This decrease in the specific angular momentum indicates a loss of the momentum as the cores evolve and become, on average, more gravitationally bound. 

\begin{figure}
\plottwo{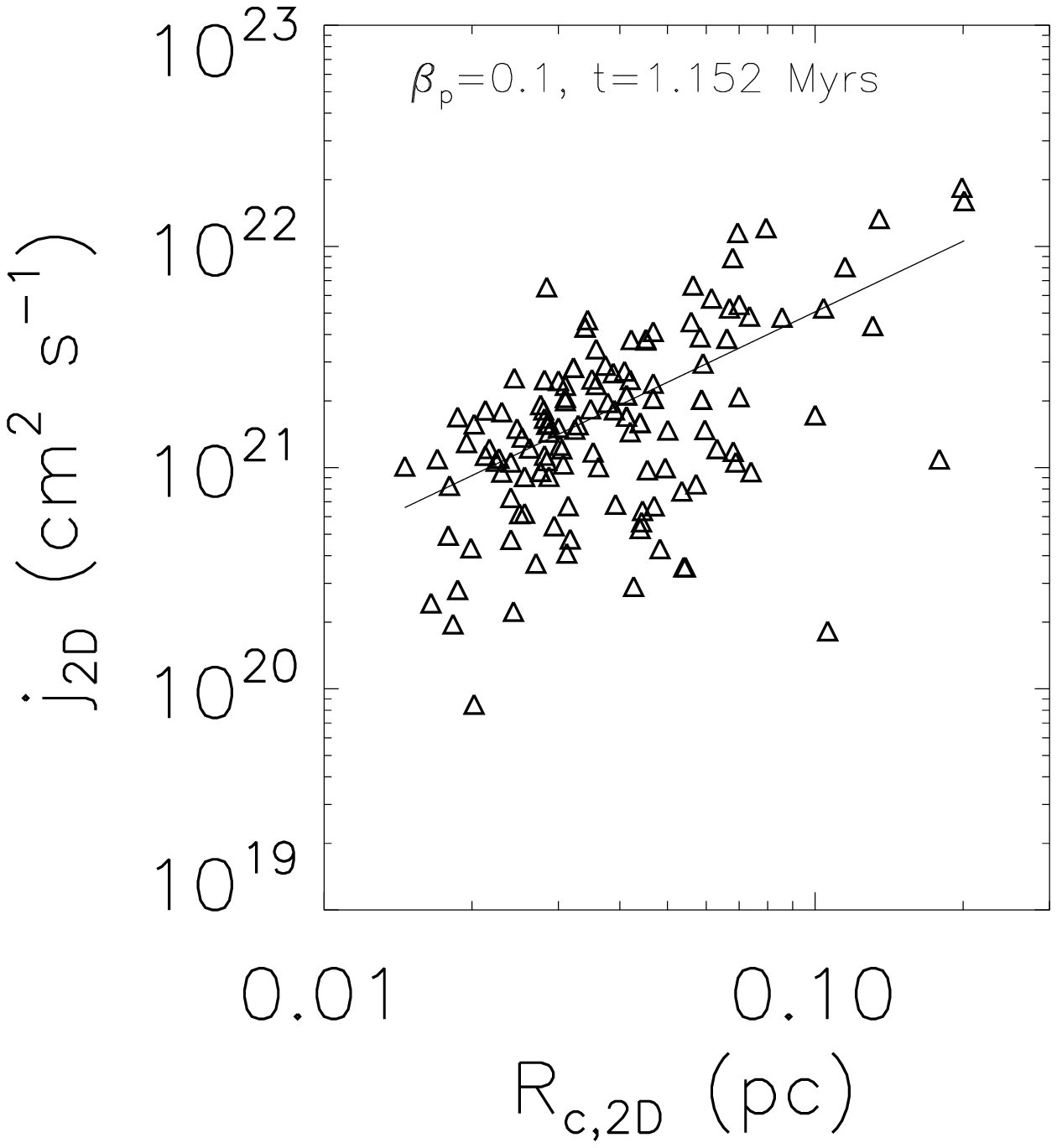}{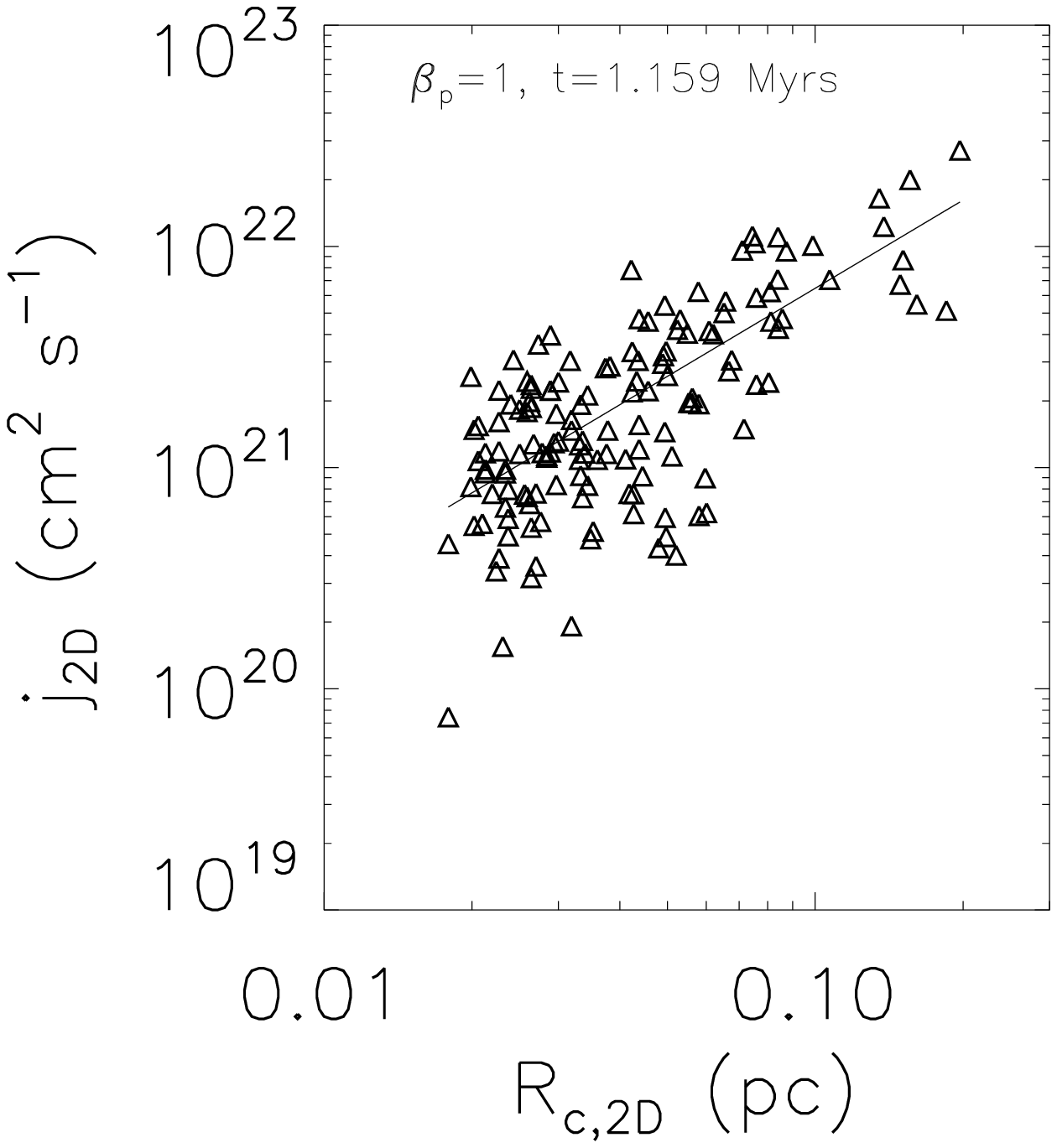}
\caption{Projected radius $R_{c,2D}$-projected specific angular momentum $j_{2D}$ relation for C160 cores identified at nearly similar epochs in simulations B1 (left) and B2 (right).}
\label{fig15}
\end{figure}

Following our earlier approach of generating observational counterparts to the relations studied in the intrinsic three-dimensional space, we construct the same $R_{2D}-j_{2D}$ relation of the cores, where $R_{2D}$ is the size of the core seen in projection and taken to be the square root of its surface and $j_{2D}$ is the same, earlier defined, specific angular momentum derived from the velocity maps of the cores. Fig.~\ref{fig15} displays the $R_{2D}-j_{2D}$ relation for the C160 cores at nearly similar epochs in the two simulations B1(left) and B2 (right). In addition to the higher normalization discussed in the previous section, the slopes of the $j_{2D}-R_{c,2D}$ relation are found to be shallower than the corresponding $j_{3D}-R_{c}$ relations observed for the same cores at the same epochs. Whereas the exponent $\lambda$ of the $j_{3D}-R_{c}$ is found to be equal to $\lambda \sim 1.8$ and $\lambda \sim 2$ at these epochs in simulations B1 and B2 respectively, the exponent of the $j_{2D}-R_{c,2D}$ relation, derived from a robust least absolute deviation fit to the data in Fig~\ref{fig15}, are $\lambda_{2D}=1.06 \pm 0.28$ and $\lambda_{2D}=1.32 \pm 0.21$ (these values are $\lambda_{2D}=1.39 \pm 0.49$ and $\lambda_{2D}=-1.28 \pm 0.11$, respectively when derived a minimization of the Chi-square error statistic method). This shows that the observational bias in the overestimate of the specific angular moment discussed in \S~\ref{distrib} affects in a more significant way the smallest cores and this results in a value of $\lambda_{2D}$ smaller than $\lambda_{3D}$. Finally, these results suggest that if a simple estimate of the specific angular momentum is to be made using a size-specific angular momentum relation, a value of $\lambda \sim 2$ should be used for the exponent of this relation rather than the smaller values derived in the observations. The derived values of $\lambda$ in the three-dimensional space which yields values of $\lambda \sim 1.8-2$ should nevertheless not be interpreted as the cores being in a state of rigid body-rotation as this is not supported by the observations of their projected velocity maps.  

\section{CONCLUSIONS}\label{conc}

In this paper, we have analyzed the rotational properties of dense molecular cloud cores formed in two magnetized, self-gravitating molecular cloud simulations with a decaying turbulence. The two simulations differ by the strength of the magnetic field in the clouds with one cloud being mildly magnetically supercritical and the other being strongly magnetically supercritical. Our results show that the formation efficiency of dense cores is strongly reduced with increasing importance of the magnetic field in the cloud (going down from 33 percent per free-fall time in the strongly supercritical cloud to 6 percent for the mildly supercritical cloud). We also observe that the median value of the specific angular momentum of the high density cores in the mildly supercritical simulation is smaller than the values derived for cores in the strongly supercritical simulation. This result is consistent with the fact that magnetic braking which leads to angular momentum loss is playing a more important role in the cloud where the magnetic field is stronger.  

We have focused our attention on the discrepancies that may arise between estimates of the specific angular momentum of the cores derived from the global velocity gradient method commonly used in the observations and its the true value measured in the intrinsic three-dimensional space. In order to derive the specific angular momentum of the cores following the observational procedure, we generate synthetic velocity maps of the cores along three different projections. The global velocity gradient of the cores is measured from the velocity map using the VFIT routine employed initially by Goodman et al. (1993). The specific angular momentum is then calculated using the global velocity gradient value under the assumption of uniform rotation of the cores. We find, in the two simulations, that the distributions of the ratio of the specific angular momentum determined in the intrinsic 3D space to the one derived from projected velocity maps peaks at values around $\sim 0.1$. This may well explain the difference by a factor $\sim 10$ that is observed between the distribution of specific angular momentum derived from the intrinsic data in our simulations and the corresponding real observations using the NH$_{3}$ and N$_{2}$H$^{+}$ molecules of roughly similar excitation density than the density thresholds used to identify the cores in the simulations.

We suggest that the origin of this discrepency (between 2D and 3D) lies in the fact that contrary to the intrinsic determination of $j$ which sums up the individual gas parcels contributions to the angular momentum, the observational determination of $j$ is based on a measurement on the global velocity gradient under the hypothesis of uniform rotation which smoothes out the complex fluctuations present in the three-dimensional velocity field. We therefore suggest that previous measurements of the specific angular momentum of the cores overestimate its true value and that a correction factor of $\sim 10$ should be applied to these measurements as well as to new determinations of the specific angular momentum when using the global gradient method adopted so far in the observations. As already stressed by other groups (e.g., Padoan et al. 1998,2000; Pichardo et al. 2000; Ostriker et al. 2001; Ballesteros-Paredes \& Mac Low 2002; Pineda et al. 2009; Federrath et al. 2010; Shetty et al. 2010) our work further highlights the importance of generating synthetic observations from three-dimensional numerical simulations that can be compared to real observations. 
    
\acknowledgements

We would like to thank the referee for her/his constructive comments. We would also like to thank Anne-Khatarina Jappsen, Stella Offner, and Chris McKee for interesting discussions on issues related to the topic of this paper. S. Dib acknowledges support from the project MAGNET of the Agence Nationale de la Recherche (France) and is very grateful to S{\o}ren Larsen for his hospitality at the Astronomical Institute in Utrecht and to Andreas Burkert for his hospitality at the Excellence Cluster Universe in Garching and to the hospitality of the Institute of Theory and Computation at Harvard University. T. Csengeri acknowledges support from the FP6 Marie-Curie Research Training Network Constellation: the origin of stellar masses (MRTN-CT-2006-035890). The numerical simulations were performed on 256 processors of the SGI ALTIX machine JADE at the Centre Informatique National de l'Enseignement Sup\'{e}rieur (CINES).  
   
{}

\clearpage 
 
\begin{deluxetable}{llllllll}
\tabletypesize{\footnotesize}
\tablecaption{Properties of the populations of C20 cores at two selected epochs in simulation B1. From left to right, the columns represent: the timesteps ($1~t_{ff,cl}=2.76$ Myrs), the sonic Mach number $\cal {M}$, the number of cores $N_{c}$, the maximum number density in the box in units of the average number density, the median mass, the median size, the median specific angular momentum, and the median rotational parameter.}
\tablewidth{0pt}
\tablehead{
\colhead{timestep} & \colhead {$\cal{M}$} & \colhead{$N_{c}$} & \colhead{$\frac{n_{max}}{n_{aver}}$}&\colhead{$M_{c,med}$ (M$_{\odot}$)} & \colhead{$R_{c,med}$ (pc)} & \colhead{$j_{3D,med}$ (cm$^{2}$ s$^{-1}$)} & \colhead{$Log_{10}(\beta_{rot,med})$}}
\startdata
$t=0.293~t_{ff,cl}$ & $2.97$ & $634$ & $1904$ & $0.0060 \pm 0.0041$& $0.021 \pm 0.011$   & $(2.52 \pm 1.18) \times 10^{20}$  & $-1.59 \pm 0.09$      \\
$t=0.424~t_{ff,cl}$ & $3.36$ & $530$ & $359319$ & $0.0045 \pm 0.0034$ & $0.019 \pm 0.010$ & $(1.76 \pm 0.81) \times 10^{20}$  &  $-1.69 \pm 0.10$ \\
\enddata
\label{table1}
\end{deluxetable}

\begin{deluxetable}{llllllll}
\tabletypesize{\footnotesize}
\tablecaption{ Similar to Tab.~\ref{table1}, but for the C20 cores of simulation B2.}
\tablewidth{0pt}
\tablehead{
\colhead{timestep} &  \colhead {$\cal{M}$} & \colhead{$N_{c}$} & \colhead{$\frac{n_{max}}{n_{aver}}$} & \colhead{$M_{c,med}$ (M$_{\odot}$)} & \colhead{$R_{c,med}$ (pc)} & \colhead{$j_{3D,med}$ (cm$^{2}$ s$^{-1}$)} & \colhead{$Log_{10}(\beta_{rot,med})$}}
\startdata
$t=0.280~t_{ff,cl}$ & $3.37$ & $640$ & $2.68 \times 10^{3}$ & $0.0053 \pm 0.0041$ & $0.020 \pm 0.010$  & $(2.02 \pm 0.99) \times 10^{20}$  & $-1.62 \pm 0.14$  \\
$t=0.420~t_{ff,cl}$ & $2.59$ &$451$ & $6.26 \times 10^{5}$ &  $0.0070 \pm 0.0052$ & $0.022 \pm 0.011$  & $(2.61 \pm 1.09) \times 10^{20}$   & $-1.85 \pm 0.08$ \\
\enddata
\label{table2}
\end{deluxetable}

\clearpage

\begin{deluxetable}{llllllll}
\tabletypesize{\footnotesize}
\tablecaption{Similar to Tab.~\ref{table1}, but for the C160 core of simulation B1.}
\tablewidth{0pt}
\tablehead{
\colhead{timestep} &  \colhead {$\cal{M}$} & \colhead{$N_{c}$} & \colhead{$\frac{n_{max}}{n_{aver}}$}&\colhead{$M_{c,med}$ (M$_{\odot}$)} & \colhead{$R_{c,med}$ (pc)} & \colhead{$j_{3D,med}$ (cm$^{2}$ s$^{-1}$)} & \colhead{$Log_{10}(\beta_{rot,med})$}}
\startdata
$t=0.293~t_{ff,cl}$ & $2.97$ & $166$ & $1.90 \times 10^{3}$ & $ 0.0061 \pm 0.0050$& $ 0.010 \pm 0.004 $   & $ (8.76 \pm 4.11 ) \times 10^{19}$  & $ -2.17 \pm 0.13 $  \\
$t=0.424~t_{ff,cl}$ & $3.36$ &$170$ & $3.59 \times 10^{5}$ & $0.0096 \pm 0.0009 $ & $ 0.012 \pm 0.006$ & $ (14.4 \pm 6.33) \times 10^{19}$  &  $ -2.19 \pm 0.14$ \\
\enddata
\label{table3}
\end{deluxetable}

\begin{deluxetable}{llllllll}
\tabletypesize{\footnotesize}
\tablecaption{ Similar to Tab.~\ref{table1}, but for the C160 cores of simulation B2.}
\tablewidth{0pt}
\tablehead{
\colhead{timestep} &  \colhead {$\cal{M}$} & \colhead{$N_{c}$} & \colhead{$\frac{n_{max}}{n_{aver}}$} & \colhead{$M_{c,med}$ (M$_{\odot}$)} & \colhead{$R_{c,med}$ (pc)} & \colhead{$j_{3D,med}$ (cm$^{2}$ s$^{-1}$)} & \colhead{$Log_{10}(\beta_{rot,med})$}}
\startdata
$t=0.280~t_{ff,cl}$ & $3.37$ & $276$ & $2.68 \times 10^{3}$ & $0.009 \pm 0.007$ & $0.011 \pm 0.005 $  & $(12.3 \pm 5.53) \times 10^{19}$  & $-2.23 \pm 0.09$ \\
$t=0.420~t_{ff,cl}$ & $2.59$ & $171$ & $6.26 \times 10^{5}$&  $0.015 \pm 0.001$ & $0.014 \pm 0.007$  & $(14.5 \pm 6.11) \times 10^{19}$   & $-2.46 \pm 0.07$ \\
\enddata
\label{table4}
\end{deluxetable}

\end{document}